\documentclass[review]{elsarticle}

\usepackage{lineno,hyperref}

\usepackage{graphicx}
\usepackage{amsmath}
\usepackage{amssymb}
\usepackage{multirow}
\usepackage{bbm}
\usepackage{comment}
\usepackage[usenames, dvipsnames]{color}

\usepackage{times}
\usepackage{subfigure}
\usepackage{nicefrac}
\usepackage{verbatim}
\usepackage{array}
\usepackage{url}
\usepackage{float}
\usepackage{multirow}
\usepackage{caption}

\usepackage{algorithmic}

\usepackage{epsfig}

\modulolinenumbers[5]

\journal{Ad Hoc Networks}

\newtheorem{thm}{Theorem}

\newdefinition{rmk}{Remark}
\newproof{pf}{Proof}
\newproof{pot}{Proof of Theorem \ref{thm2}}

\begin{document}

\begin{frontmatter}

\title{On Optimal Resource Allocation in Virtual Sensor Networks}

\author[mymainaddress]{Carmen Delgado}

\author[mymainaddress]{Jos\'e Ram\'on G\'allego\corref{mycorrespondingauthor}}
\cortext[mycorrespondingauthor]{Corresponding author}
\ead{jrgalleg@unizar.es}

\author[mymainaddress]{Mar\'ia Canales}

\author[mysecondaryaddress]{Jorge Ort\'in}

\author[mytertiaryaddress]{Sonda Bousnina}

\author[mytertiaryaddress]{Matteo Cesana}

\address[mymainaddress]{Arag\'on Institute of Engineering Research, Universidad de Zaragoza, Spain}
\address[mysecondaryaddress]{Centro Universitario de la Defensa, Zaragoza, Spain}
\address[mytertiaryaddress]{Dipartimento di Elettronica, Informazione e Bioingegneria, Politecnico di Milano, Milano, Italy.}

\begin{abstract}
Sensor network virtualization is a promising paradigm to move away from highly-customized, application-specific wireless sensor networks deployment by opening up to the possibility of dynamically assigning general purpose physical resources to multiple stakeholder applications. In this field, this paper introduces an optimization  framework to perform the allocation of physical shared resources of wireless sensor networks to multiple requesting applications. The proposed optimization framework aims at maximizing the total number of applications which can share a common physical network, while accounting for the distinguishing characteristics and limitations of the wireless sensor environment (limited storage, limited processing power, limited bandwidth, tight energy consumption requirements). Due to the complexity of the optimization problem, a heuristic algorithm is also proposed. The proposed framework is finally applied to realistic network topologies to provide a detailed performance evaluation and to assess the gain involved in letting multiple applications share a common physical network with respect to one-application, one-network vertical design approaches.  
\end{abstract}


\end{frontmatter}


\section{Introduction}

In the Internet of Things (IoT) vision, the Internet is ``pushed down" to everyday objects which are equipped with sensing capabilities to gather information on the environment they are immersed in, processing/storage capabilities to locally filter and store date, and communication peripherals to deliver the collected/processed data remotely either directly, or through multi-hop paths leveraging the cooperation of other smart objects for traffic relaying.  In this last case, network of smart objects, often referred to as Wireless Sensor Networks, are set up to collect and deliver data in specific areas.   WSNs can be deployed in diverse scenarios and environments to support diverse application/services ranging from smart home or environmental monitoring based on scalar sensed data to more demanding applications based on multimedia sensors.   

Usually, WSNs are designed and deployed  in a  ``vertical'', application-specific way, in which the hardware and network resources are customized to the specific application requirements. On one hand, such design paradigm allows to have ``optimal'' performance on the specific application, but, on the other hand, it precludes resources (hardware and software) reuse when other applications and services must be contemplated. In the end, this has led in the past to the proliferation of redundant WSNs deployments \cite{Jayasumana07}.

In this context, novel approaches are recently being investigated targeting the smart reuse of general purpose wireless sensor networks to dynamically support multiple applications and services. The key idea behind these approaches, which often go under the names of Virtual Sensor Networks (VSN) \cite{Islam12b} or Software Defined Sensor Networks (SDSN) \cite{luo2012CL}, is to decouple the physical infrastructure and resources from application ownerships which leads to  more efficient resource utilization, lower cost and increased flexibility and manageability in WSN deployments \cite{Sarakis12}. Network virtualization technologies are used to abstract away ``physical  resources'' including node processing/storage capabilities, available communication bandwidth and routing protocols, which can then be ``composed'' at a logical level to support usage by multiple independent users and even by multiple concurrent applications \cite{Merentitis13}. 
While network virtualization and software defined networks are already a reality in many communication networks \cite{Chowdhury10, Xiao13},  research on sensor network virtualization is still in its infancy and comprehensive solutions   still need to be found to cope with the specific characteristics of WSNs in terms of limited node capabilities and communication bandwidth. 

In this work, we focus on the design of a virtualization engine for WSNs. Namely, we consider a general purpose WSN which can be used to support multiple applications and we propose a mathematical programming framework to optimally allocate shared physical resources to the requesting applications. In more details, the proposed framework allocates the physical resources of the general purpose WSN to multiple concurrent applications while accounting for the network- and hardware-specific constraints  (processing, storage, available bandwidth, limited communication range) and the specific application requirements. Due to the high computational complexity of the resulting optimization model, a heuristic algorithm is also proposed. Numerical results are then obtained by applying the proposed framework to realistic WSN instances to assess the efficiency of the virtualization process. 

The paper is organized as follows: Section \ref{sec:related} overviews the related work in the field of sensor network virtualization. Section~\ref{model} describes the proposed system model and the optimization problem for resource allocation in virtual sensor networks, including a complexity analysis of this problem.  Section~\ref{heuristic} explains the proposed heuristic algorithm. In Section~\ref{performance}, the proposed optimization model and heuristic algorithm are evaluated by simulation for a set of scalar and multimedia applications also with different types of sensor nodes. Finally, some conclusions are provided in Section~\ref{conclusion}. 

\section{Related Work}
\label{sec:related}
The emergence of shared sensor networks has stimulated research efforts in the field of novel programming abstractions at the node level and management framework at the network level to support multiple applications over a shared physical infrastructure \cite{Fischer13},\cite{Islam12}, \cite{Madria14}, \cite{Khan15}.  

At the node level, architectures based on virtual machines are proposed to enable virtualization and re-programmability. As an example, Mat\'e \cite{Levis2002}, ASVM \cite{Levis2005}, Melete \cite{Yu2006} and VMStar \cite{Koshy2005} are  frameworks for building application-specific virtual machines over constrained sensor platforms. 

At the network level, several virtualization management platforms have been introduced. SenSHare \cite{Leontiadis2012} creates multiple overlay sensor networks which are ``owned'' by different applications on top of a shared physical infrastructure. UMADE \cite{5465985} is an integrated system for allocating and deploying applications in shared sensor networks based on the concept of Quality of Monitoring (QoM). Fok \emph{et al.} \cite{Fok2011} introduce middleware abstractions to represent multiple QoM requirements from multiple applications, whereas a service-oriented middleware is presented in \cite{Li20141775} to address the challenges faced by running multiple applications onto heterogeneous WSNs. A prototype of Software Defined Wireless Sensor Network is proposed in \cite{Huang2015IJDSN} where a centralized control plane dynamically manages communication routes in the network with the goal of augmenting the energy efficiency.  

Generally speaking, the aforementioned work provides ``practical'' building blocks to build up virtual sensor networks. Differently, we focus in this paper on the ``intelligence'' to properly allocate physical resources to virtual applications, which can be cast as a general resource allocation problem. Even if radio/network resource allocation is a widely debated topic in the literature, still very few works have appeared on the optimal resource allocation in the field of Virtual or Shared Sensor Networks. 

In \cite{Xu2010} the authors propose an optimization framework to maximize the  Quality of Monitoring (QoM) in shared sensor networks. The proposed framework focuses on environmental monitoring applications whose reference ``quality'' can be modeled as dependent on the variance in the sensed data, and derives the application-to-sensors assignment which minimizes such variance. The same authors address in a later work the case where the application assignment problem is no longer centralized but rather distributed by resorting to game-theoretic tools \cite{6195490}.  Ajmal \emph{et al.} leverage the concept of QoM and propose an admission control scheme to dynamically ``admit'' applications to be deployed on physical sensor networks. The authors of \cite{6496703}  focus on the problem of scheduling applications to shared sensor nodes with the ultimate goal of maximizing the sensor network lifetime.   Along the same lines, Zeng \emph{et al. } propose in \cite{Zeng2015TC} an optimization framework to prolong network lifetime by properly scheduling the tasks in a shared/virtual sensor network. 

The problem of allocating physical resources to multiple application is also often also cast as an auction.  In \cite{SEC:SEC631}, the authors propose a reverse combinatorial auction, in which the sensor nodes act as bidders and bid cost values (according to their available resources) for accomplishing the subset of the applications' tasks. Optimal bidding strategies are then studied to make the auction effective and truthful. 

This work extends our previous work in \cite{Delgado2015}, where the preliminary optimization framework to allocate resources in virtual sensor networks is introduced; building on the aforementioned work, we further provide here a complexity analysis of the proposed optimization problem which is proven to be NP-hard, and we propose a heuristic iterative algorithm to obtain sub-optimal solutions of this resource allocation problem in reduced computation time. Finally, we provide here a comprehensive performance evaluation of the proposed approach: we analyze the impact of varying the main model parameters (number of scalar and multimedia nodes; number of sinks; lifetime; type of routing) in the system performance and we also evaluate the performance of the proposed heuristic algorithm.

\section{System model and optimization framework}
\label{model}

Let $S = \left\{s_1,s_2, \dotsc , s_l\right\} $ be a set of sensor nodes, $A = \left\{a_1,a_2, \dotsc , a_m\right\}$ a set of applications which are to be deployed in the reference area, and $T = \left\{t_1, \dotsc ,t_n \right\}$ a set of test points in the reference network scenario. To simplify notation, in the following we will use the subscript index $i$ (or $h$) to refer to a sensor node $s_i$ (or $s_h$), the subscript index $j$ to refer to an application $a_j$ and the subscript index $k$ to refer to a test point $t_k$.

Each application $j$ requires to cover a given set of test points $T_j \subseteq T$. Formally, the application $j$ has to be deployed on a subset of $S$ such that all the test points in $T_j$ are covered. A test point is covered by a sensor node $i$ if it is within its sensing range, $R_i^{s}$. It is convenient to introduce as well the set $S_{jk}$ defined as the set of sensor nodes which physically cover the test point $k$, with $k \in T_j$. In other words, if the application $j$ is deployed on any of the sensors in set $S_{jk}$, then the target test point $k$ is covered for this application. A necessary condition for an application $j$ to be successfully deployed is that all the test points in its target set $T_j$ must be covered. 

Each application $j$ in $A$ is further characterized by a requirement vector $r_j = \left\{c_j,m_j,l_j\right\}$ which specifies the required source rate [bit/s], memory [bits] and processing load [MIPS] consumed by the application when it is deployed on a sensor node. The requirement vector can be interpreted as the amount of resources needed to accomplish the specific tasks required by the application (e.g., acquire, process and transmit 10 temperature samples, or acquire process and transmit one JPEG image, etc.). Additionally, each sensor node $i$ in $S$ is characterized by a given resource vector $o_i = \left\{C_i, M_i, L_i, E_i\right\}$, which specifies its available bandwidth, storage capabilities, processing power and energy store. 

A protocol interference model with power control \cite{Shi13} is used to characterize the wireless communications among the sensor nodes. The maximum transmission power is $P_{max}$. With this power, there are a maximum transmission range $R^T_{max}$ and a maximum  interference range $R^I_{max}$. Given a directional link between a pair of nodes $\left(i,h\right)$, the channel gain from transmitter $i$ to receiver $h$ is defined as $g_{ih} = g_0 \cdot d_{ih}^{-\gamma}$, being $d_{ih}$ the distance from $i$ to $h$, $\gamma$ the path loss index and $g_0$ a constant dependent on antenna parameters. A transmission is successful if the received power exceeds a threshold $\alpha$. Additionally, all the nodes under the interference range of a sensor node share the same transmission channel and therefore, the transmission time must be divided between them. If $p_i$ is the transmission power assigned to node $i$, a transmission towards $h$ is successful if $p_i \cdot d_{ih} > \alpha$. Thus, the transmission range for node $i$ with transmission power $p_i$ can be obtained as $R_i^T\left(p_i\right)=\left(p_i \cdot g_0/\alpha\right)^{-\gamma}$. Similarly, the interference resulting from node $i$ with power $p_i$  is non-negligible only if it exceeds a certain threshold $\beta$. Then, the interference range is $R_i^I\left(p_i\right)=\left(p_i \cdot g_0/\beta\right)^{-\gamma}$.

Qualitatively, the application assignment problem for virtual sensor networks can be defined as follows: to maximize the weighted number of deployed applications subject to coverage constraints (the set of test points of each application must be covered) and application requirements (each application should be assigned enough bandwidth, and processing and storage resources to operate successfully). In addition, due to the multihop nature of WSNs, routing and link capacity constraints must be considered when the data generated by the application has to be delivered remotely.

Further, let us assume that a preference vector across all the $m$ applications is defined $Q = \left[q_1, q_2,\dotsc ,q_m\right]^T$ where $q_j$ represents the revenue for the network provider for having application $j$ successfully deployed in the network. Let $z_j$ be a binary variable indicating if application $j$ is successfully deployed in the network. Let $y_{ijk}$ be a binary variable indicating if test point $k$ of application $j$ is deployed at sensor node $i$. Let $x_i$ be a binary variable indicating if sensor node $i$ is active in the network. Let $h_{jk}$ be a binary variable which indicates if test point $k$ belonging to set $T_j$ is actually covered by a sensor node which runs application $j$. 

The objective function aims at maximizing the overall revenue out of the application deployment process while minimizing the cost related to activating sensor nodes:

\begin{equation}
\label{eq:maximiza}
\max \left( \sum_{j \in A} q_j  z_j - \sum_{i \in S} \delta_i x_i  \right)
\end{equation}

\noindent where $\delta_i$ is the cost incurred in activating sensor node $i$.

\subsection{Constraints on coverage and on resources of the sensors}
\label{coverage-const}

Constraints (\ref{eq:const2})-(\ref{eq:const3}) require that all the applications which are actually deployed do fulfill the coverage constraints, that is, they cover all the required test points. Specifically, Eq. (\ref{eq:const2}) indicates if test point $k$ of an application $j$ is effectively covered. If so, it ensures that it is covered by only one sensor node $i$. Eq. (\ref{eq:const2b}) ensures that if a sensor $i$ does not cover a test point $k$ of an application $j$, then it will not sense that test point. Depending on the application, it can be possible that the same sensor node can cover several of its test points (e.g., visual applications). If we define $N_{ij}$ as the maximum number of test points of the same application $j$ that a sensor $i$ is able to cover, Eq. (\ref{eq:const21}) guarantees that this threshold is not exceeded. Eq. (\ref{eq:const3}) indicates that if an application $j$ is successfully deployed, i.e., if $z_j = 1$, then all the test points of application $j$ must be covered. In addition, it guarantees that if the application cannot be deployed, none of its test points are covered so that no resources are wasted.  
 
\begin{gather}
\sum_{i \in S_{jk}} y_{ijk} = h_{jk}  \qquad  \forall j \in A, \forall k \in T_j  \label{eq:const2} \\
y_{ijk} = 0  \qquad  \forall i \notin S_{jk}, \forall j \in A, \forall k \in T_j  \label{eq:const2b} \\
\sum_{k \in T_j} y_{ijk} \leq N_{ij}  \qquad \forall i \in S,  \forall j \in A  \label{eq:const21} \\
z_j =  \frac{ \sum_{k \in T_j} h_{jk}} {\left| T_j \right|}   \qquad   \forall j \in A \label{eq:const3} 
\end{gather}

Constraints (\ref{eq:const5}) and (\ref{eq:const6}) are budget-type constraints for the available storage and processing load of the sensor nodes. 

\begin{gather}
\sum_{j \in A} \sum_{ k \in T_j} m_j y_{ijk} \leq M_i  \qquad   \forall i \in S  \label{eq:const5} \\
\sum_{j \in A} \sum_{ k \in T_j} l_j y_{ijk} \leq L_i  \qquad   \forall i \in S  \label{eq:const6}
\end{gather}

\subsection{Routing constraints}
\label{routing-const}

The deployed applications will require most likely that the information generated locally is delivered remotely to collection points (sink nodes) through multihop paths. Note that these sensor nodes may run deployed applications or not. By resorting to a fluid model, it should be ensured that all the data produced by the sensors running applications are received by the sink nodes. This fact can be conveniently expressed using the following constraints:

\begin{equation}
\label{eq:const82}
\sum_{\substack{ h \in S \\ i \neq h }}f_{hi} - \sum_{\substack{h \in S \\ h \neq i} }f_{ih} + \sum_{j \in A}\sum_{ k \in T_j } c_j y_{ijk} = 0   \quad   \forall i \in S \setminus SINK 
\end{equation}

\begin{equation}
\label{eq:const103}
\sum_{j \in A} |T_j|  c_j z_j = \sum_{ h \in SINK } \left(\sum_{\substack{i \in S \\ i \neq h}} f_{ih}   + \sum_{j \in A} \sum_{ k \in T_j} c_j y_{hjk} \right)
\end{equation}

\noindent where $SINK$ is the set of sink nodes (a subset of $S$) and $f_{ih}$ is a variable representing the flow of data in bps transmitted from node $i$ to node $h$. Constraints (\ref{eq:const82}) enforce flow conservation at  sensor nodes. 
Constraint (\ref{eq:const103}) imposes that all the generated data are delivered to the set of sinks. The last term of this expression allows the sinks to be running applications as well. 

The following constraint set enforces that if a sensor node is either running an application or receiving data, then it must be active in the network:

\begin{equation}
\label{eq:const9}
\sum_{\substack{h \in S \\ h \neq i}}f_{hi} + \sum_{j \in A}\sum_{ k \in T_j } c_j y_{ijk} \leq K x_i  \qquad   \forall i \in S  
\end{equation}

\noindent where $K$ is a constant high enough (higher than the maximum transmission rate of a node). Finally, constraints

\begin{equation}
\label{eq:const81}
f_{ih} \leq K l_{ih}    \qquad   \forall i, h \in S  
\end{equation}

\noindent where $l_{ih}$ is a constant that indicates if there is a viable link between $i$ and $h$, i.e., if the distance between both nodes is less than the maximum transmission range $R^T_{max}$, then $l_{ih} = 1$ and $l_{ih} = 0$ otherwise. Therefore, these constraints ensure that data must be transmitted exclusively along neighboring nodes.

These expressions allow flow splitting and multipath routing. In the sequel, we will denote this kind of routing as \textit{multipath routing}. 

However, in WSNs routes from each sensor node to a sink node follow typically a single path, such as the Destination Oriented Directed Acyclic Graph (DODAG) of RPL \cite{rfc6550}. Therefore, we introduce the following restrictions to ensure that all the traffic flowing out of a sensor has only one possible route to a sink:

\begin{gather}
g_{ih} \leq l_{ih}   \qquad   \forall i,h \in S  \label{eq:constf11} \\
\sum_{h \epsilon S} g_{ih} \leq 1   \qquad   \forall i \in S  \label{eq:constf12} \\
f_{ih } \leq K g_{ih}    \qquad   \forall i,h \in S  \label{eq:constf13}
\end{gather}

\noindent where $g_{ih}$ is a binary variable which indicates if data are transmitted from node $i$ to node $h$. Constraints (\ref{eq:constf12}) and (\ref{eq:constf13}) impose that only one link from sensor node $i$ to any of its neighbors transports all the data that $i$ must forward. In the sequel, we will denote this kind of routing as \textit{singlepath routing}.

Including the route creation in the optimization framework may not be always feasible. In addition, since all the traffic in WSN is forwarded to a single or a limited number of sinks, the main bottleneck will be mainly the last hop to these sinks. For these reasons, we also consider the possibility of excluding the routing from the optimization process and assuming a predefined set of routes from each node to a sink. To that purpose, we build DODAGs using the number of hops as a metric (i.e., when there are several sinks, each node belongs to the DODAG that reaches a sink with the minimum number of hops). This implies that equations (\ref{eq:constf11})-(\ref{eq:constf13}) must be excluded from the model and that for each node $i$, the constant $l_{ih}$ is 1 just for a single $h$ (the father node in the routing tree towards the sink) and therefore $f_{ih'} = 0$ for all $h' \neq h$. In the sequel, we will denote this kind of routing as \textit{static routing}.

\subsection{Bandwidth constraints}
\label{bandwidth-const}

The available bandwidth in the network is limited and must be shared among sensor nodes. We assume that a fair medium access control schemes orchestrate the access to the shared medium. Given a directional link between a pair of nodes $\left(i,h\right)$, let the capacity of the link be defined as $C_{ih} = \min\left(C_i, C_h\right)$. This aims to model that the transmission rate is limited by the most restrictive node in the link. Transmissions of other links where $i$ or $h$ are either transmitter or receiver cannot be simultaneously active with $\left(i,h\right)$ (note that some combinations are not possible in this particular case due to routing constraints, i.e., another link with $i$ as a transmitter). 

According to the considered protocol interference model, the interfering links for link $\left(i,h\right)$ are those whose receiver is within the interference range of node $i$ or the links where $j$ is within the interference range of its transmitter. Although none of these links can be simultaneously active with $\left(i,h\right)$, some of them (depending on their relative positions) could be simultaneously active with each other. Therefore, if we define $IF_{ih}$ as the fraction of time that other links interfere the link $\left(i,h\right)$, we have that: 

\begin{align}
\label{eq:constf120}
IF_{ih} = \sum_{\substack{g \in S \\ g \neq h}} \frac{f_{ig}}{C_{ig}} +  \sum_{g \in S} \frac{f_{gi}}{C_{gi}} + \sum_{\substack{g \in S \\ g\neq i}} \frac{f_{hg}}{C_{hg}} +  \sum_{\substack{g \in S \\ g\neq i}} \frac{f_{gh}}{C_{gh}} + \nonumber \\   \sum_{\substack {g,t \in S \\ d_{it}<R_i^I(p_i) }} \frac{f_{gt}}{C_{gt}} + \sum_{\substack {g,t \in S \\ d_{gh}<R_g^I(p_g) }} \frac{f_{gt}}{C_{gt}} 
\end{align}

Then, for each link $\left(i,h\right)$ in the network it must be ensured that the fraction of time used by the link plus all its interferences is less or equal to 1: 

\begin{equation}
\label{eq:constf240}
\frac{f_{ih}}{C_{ih}} + IF_{ih} \leq 1  \qquad   \forall i,h \in S  
\end{equation}

Constraints (\ref{eq:constf240}) are the equivalent budget-type constraints for the available wireless capacity to the storage and processing load constraints given in (\ref{eq:const5}) and (\ref{eq:const6}). 

\subsection{Energy constraints}
\label{energy-const}

Finally, energy constraints are included to ensure that the application deployment pattern guarantees a minimum lifetime $L$ for the virtual sensor network. Typically, energy consumption due to wireless communication (i.e., transmitting and receiving) has been considered the dominant factor in power consumption for WSNs \cite{Akyildiz02}. While this is the case for traditional scalar applications, where processing is limited to simple operations, in multimedia applications the energy required to process data can not be neglected \cite{Redondi12}.

Regarding wireless transceiver, the power dissipation at the radio transmitter $P^{t}_{i}$ or at the radio receiver $P^{r}_{i}$ of each node $i$ can be modeled as \cite{Hou08}:

\begin{gather}
\label{eq:powertxall}
P^{t}_{i} = \sum_{h \in S, h \neq i} \left(\beta_1 + \beta_2  d^{\gamma}_{ih} \right)  f_{ih}   \qquad   \forall i \in S \\
\label{eq:powerrxall}
P^{r}_{i} = \rho  \sum_{h \in S, h \neq i} f_{hi}   \qquad   \forall i \in S 
\end{gather}

Typical values for $\beta_1$, $\beta_2$ and $\rho$ are $ \beta_1 =\rho = 50$ nJ/bit and  $\beta_2 = 0.0013 \text{pJ/bit/m}^4$,  with $\gamma = 4$ the path loss index. 

The estimation of the power dissipation due to the processing load is not so straightforward, since it depends on several factors such as the hardware architecture of the nodes or the specific implementation of the algorithm for each application. In the lifetime constraints set in (\ref{eq:constf110}), this power dissipation is left as a function $f$ of the processing loads $l_j$ of the applications. In Section~\ref{performance}, further details about the specific evaluated multimedia applications are given.

\begin{equation}
\label{eq:constf110}
P^{t}_{i} + P^{r}_{i} + f\left(\sum_{j \in A} \sum_{ k \in T_j} y_{ijk}  l_j\right) \leq \frac{E_i}{L}
 \qquad   \forall i \in S  
\end{equation}

\subsection{Complexity analysis}\label{complexity}
\begin{thm}
\label{thm1}
The application deployment problem is NP-complete.
\end{thm}
\begin{pf}
The NP-completeness can be proved by restriction, that is by showing that an NP-complete problem reduces to our application deployment problem. The reference problem we use in the proof is the multiple Knapsack Problem which is known to be NP-complete. 
Let's consider the particular instance of the application deployment problem characterized by the following setting: $\delta_i=$0,$\forall i \in S$ (negligible sensor activation cost), and $S_{jk}=S$, $\forall j \in A$, $T_j=\{1\}$, $\forall j \in A$ (all the applications need to cover one single test point which is reachable from all the sensor nodes). 
Let's further assume that routing is not needed, that is, formally $SINK=S$, and that sensor nodes do not have processing capability limitation, $L_i = \infty$, $\forall i \in S$. In such setting, since all the applications need to cover one single test point which is "reachable" from all the sensor nodes (from $T_j=\{1\}$ and $S_{jk}=S$), the index $k$ can be safely dropped from variables $y_{ikj}$ and $h_{jk}$.
The application deployment problem can be re-written as follows: 
\begin{equation}
\max \sum_{j \in A} q_j  z_j \nonumber
\end{equation}
s.t.
\begin{align*}
\sum_{i \in S} y_{ij} = h_{j}  & \qquad  \forall j \in A  \\\nonumber
y_{ij} \leq 1  & \qquad \forall i \in S,  \forall j \in A  \\\nonumber
z_j =   h_{j}   & \qquad   \forall j \in A  \\\nonumber
\sum_{j \in A} m_j y_{ij} \leq M_i  &\qquad   \forall i \in S \nonumber
\end{align*}
which can be further simplified as:
\begin{equation}
\max \sum_{j \in A} \sum_{i \in S} y_{ij}  q_j \nonumber
\end{equation}
\begin{align*}
\sum_{i \in S} y_{ij} \leq 1  &\qquad \forall  j \in A  \\\nonumber
\sum_{j \in A} m_j y_{ij} \leq M_i & \qquad   \forall i \in S \nonumber
\end{align*}
This last formulation is a multi-knapsack problem which is known to be NP-complete. By restriction, also the application deployment problem must be NP-complete \cite{Garey}.$\square$ 
\end{pf}

\section{Heuristic algorithm}
\label{heuristic}

\begin{figure*}
		\centering
		{
			{\includegraphics[width=4in]{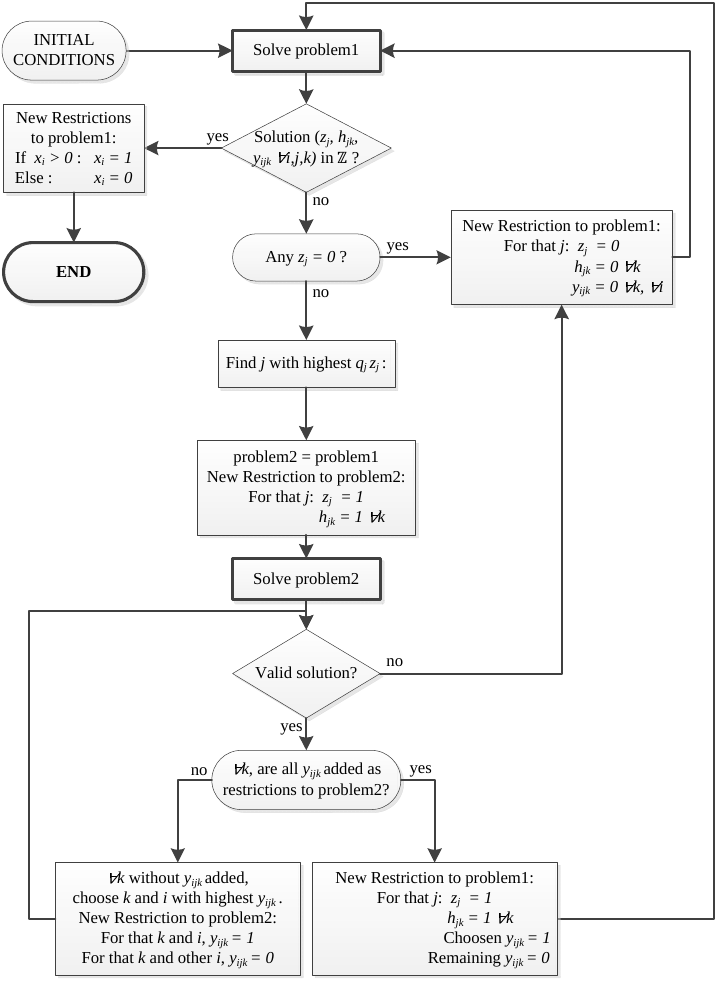}}			
			\caption{Diagram of the heuristic algorithm}
			\label{fig:Heuristicdiagram}			
		}
\end{figure*}

In order to obtain sub-optimal solutions of the resource allocation problem in reduced computation time, we introduce here a heuristic iterative algorithm which is based on LP relaxation the original problem. In short, we drop the integrality constraints on variables $z_j$, $x_i$, $y_{ijk}$ and $h_{jk}$ and focus on the \textit{static routing} strategy defined in section~\ref{routing-const}, and iteratively solve simplified problems further checking feasibility of the obtained solution in the original problem at each iteration step. Fig.~\ref{fig:Heuristicdiagram} shows the diagram block of the algorithm, whose steps are explained next.

First, the relaxed LP problem is solved (\emph{problem1}). If all the variables in the obtained solution are integers, then the solution is also valid for the original MILP problem and the algorithm terminates. If not, we see if there is any application that is not active at all ($z_j=0$) in the solution of the relaxed problem and we add new restrictions to \emph{problem1} forcing this application to be inactive (if there are more than one inactive applications, we choose one at random). We follow this procedure of solving \emph{problem1} and adding these restrictions until $z_j>0$ for all the remaining applications.

Then, we choose the application $j$ with maximum value of $q_j z_j$ and we try to activate it when solving the optimization problem. To that purpose, we define a new temporary problem (\emph{problem2}) equal to \emph{problem1}, but with additional restrictions to force application $j$ to be active. If \emph{problem2} does not have a feasible solution, we dismiss application $j$ and we add permanent restrictions to \emph{problem1} forcing this application to be inactive. On the other hand, if \emph{problem2} has a valid solution, then the values of $y_{ijk}$ still have to be obtained: The feasibility of the solution guarantees that application $j$ is active ($z_j=1$) and therefore that all its test points are covered ($h_{jk}=1$). However, it does not guarantee that each test point $k$ is covered by a single node $i$ ($y_{ijk}=1$) as the variables $y_{ijk}$ are still relaxed.

To solve this problem we proceed as follows: for each test point $k$ we choose the node $i$ with the highest value of $y_{ijk}$ and we add to \emph{problem2} the restriction $y_{ijk} = 1$ to force node $i$ to cover completely test point $k$. If \emph{problem2} with these new restrictions is feasible, then we add all the restrictions to \emph{problem1} to guarantee that application $j$ is active. On the contrary, if \emph{problem2} does not have a valid solution, then application $j$ is dismissed. Once the solution of \textit{problem1} fulfills the integrality of variables $z_j$, $h_{jk}$ and $y_{ijk}$, then we set $x_i = 1$ if $x_i>0$ and $0$ otherwise.

\section{Performance evaluation}
\label{performance}

The proposed model leads to a mixed integer linear programming (MILP) problem, which has been solved using the CPLEX software \cite{cplex}. To evaluate the model we have considered a scenario with two different types of sensor nodes and four different applications (two scalar and two multimedia). Next we define the main features of both sensor nodes and applications and the simulation parameters. Then, results are presented.

\subsection{Sensor nodes}
\label{sensornodes}

We have considered TelosB sensor motes \cite{TelosB} and BeagleBone nodes \cite{beagle}. The energy budget for both nodes is 32400 J assuming that a node runs at 3 V with 3 Ah of battery supply (2 AA batteries). Each TelosB mote has integrated a temperature and a light sensor, an IEEE 802.15.4 radio with integrated antenna and a 8 MHz TI MSP430 microcontroller which can operate at 8 MIPS and with 10 KB RAM, although only 7 KB are available for applications \cite{5465985}. Therefore, their resource vector is $o_i = \left\{C_i, M_i, L_i, E_i\right\} = \left\{\text{250 kbps, 7 KB, 8 MIPS, 32400 J}\right\}$.  These motes are suitable for supporting scalar applications. BeagleBone is a low-power platform based on a Linux Computer that includes 720 MHz super-scalar ARM Cortex-A8 processor (up to 720 MIPS) and 256 MB of RAM. BeagleBone nodes should include a Shimmer Span IEEE 802.15.4-compliant transceiver, a low-power USB camera for multimedia applications and also scalar sensors. Their resource vector is $o_i = \left\{\text{250 kbps, 256 MB, 720 MIPS, 32400 J}\right\}$.  


	\subsection{Applications}

For the scalar applications, we have considered temperature and light monitoring. Temperature monitoring applications require 4462 bytes of RAM, while light monitoring applications require 1006 bytes \cite{5465985}. This kind of applications has a low sample rate (0.017-1 Hz according to \cite{MultimediaData}). We have chosen a sample rate of 0.5 Hz for temperature monitoring and 1 Hz for light monitoring. Considering a packet size of 127 bytes per sample, the temperature application has a source rate of 0.5 kbps, whereas for the light application is 1 kbps. Given these parameters, we can assume that the processing load $l_j$ is negligible in the application requirement vector, (i.e. memory usage, transmission rate or the energy consumed by the transmission will be more limiting factors than the processing load or the energy consumed by the processing). Thus, the requirement vector for temperature monitoring is $r_j = \left\{c_j,m_j,l_j\right\} = \left\{\text{ 0.5 kbps, 4462 B, -}\right\}$
whereas for light monitoring is $r_j = \left\{\text {1 kbps, 1006 B, -}\right\}$

For multimedia applications we focus on visual sensor networks, i.e. WSNs designed to perform visual analysis (e.g. object recognition) \cite{Redondi12}. We consider two paradigms, the classic Compress-Then-Analyze (CTA) and the opposite approach, Analyze-Then-Compress (ATC) \cite{Redondi12}, \cite{Thesis}. CTA applications are those where images acquired from camera nodes are compressed and sent to a central controller for further analysis. On the other hand, ATC applications are those where camera nodes perform visual feature extraction and transmit a compressed version of these features to a central controller. In \cite{Thesis} a detailed characterization of transmission rates and energy consumption for both approaches is provided. In order to evaluate the model with some realistic parameters, we have chosen some specific values for both cases based on the aforementioned study. It is assumed that different techniques for the extraction of local visual features are used: CTA will use the SIFT (Scale Invariant Feature Transform) algorithm while ATC will use BRISK (Binary Robust Invariant Scalable Keypoints) algorithm. Assuming a desired Mean of Average Precision (MAP) of 0.6, the use of Zurich Building Database (ZuBuD) \cite{ZuBuD} and an application frame rate of  $\lambda = 1$ query per second for both CTA and ATC paradigms, the needed capacity will be 20 kbps for CTA-SIFT and 12 kbps for ATC-BRISK \cite{Thesis}.

For this kind of applications, the energy consumed to process the data is not negligible. In~\cite{Thesis} a characterization of this energy on a BeagleBone-based visual sensor node is provided. The processing energy for the CTA paradigm can be computed as:
 \begin{equation}
\label{eq:EcpuCTA}
 E_{cpu}^{CTA} (\rho) = P_{cpu} \cdot t_{cpu}^{CTA} \left(\rho \right)  
\end{equation}
 
\noindent where $P_{cpu}$ is the power dissipated by the processor of the visual sensor node and has a value of 2.1 W for BeagleBone sensor nodes; and $t_{cpu}^{CTA} \left(\rho \right)$ is the time required to process an image, which depends on $\rho$, the amount of sent information per query (20 kbs in our scenario). According to the results in \cite{Thesis}, the processing energy for an image for the CTA application in our scenario is 0.05 J. Therefore, assuming a frame rate of $\lambda = 1$ query per second, the power dissipation (function $f$ in eq.  (\ref{eq:constf110})) is 0.05 W. In addition, we can estimate the required processing load $l_j$ for a BeagleBone as the fraction of time used by the application ($t_{cpu}^{CTA} \cdot \lambda$) multiplied by the processing power of the sensor node, $L_i$. In this case, $24.5 \cdot 1\cdot 720 =17.64$ MIPS.

Similarly, the processing energy for the ATC paradigm can be computed as:
 \begin{equation}
\label{eq:EcpuATC}
 E_{cpu}^{ATC} (\rho) = P_{cpu} \cdot \left[ \tau_{off} + M(\rho) \cdot \left(\tau_{det} + \tau_{desc}\right) \right] 
\end{equation}
 
\noindent where $\tau_{off}$ is the time spent for initializing the detector and has a value of $1.6 \cdot 10^{-4}$ ms/pixel. With an image size of $640\times480$ pixels, $\tau_{off}$ is 49.152 ms. $\tau_{det}$ and $\tau_{desc}$ are the time spent for detecting and describing one BRISK feature of the image and their values are 0.31 ms and 0.16 ms respectively. $M(\rho)$ is the optimal number of features that depends on the rate $\rho$. For $\rho = 12$ kb/query, the minimum value of $M$ to provide a MAP of 0.6 is $M = 100$ features. Thus the processing energy for an image for the ATC application in our scenario is 0.2 J, and the power dissipation is 0.2 W. The processing load in this case is 69.23 MIPS.

Regarding memory requirements, specific values have not been obtained for these applications. However, given the great difference in the amount of available memory in TelosB (10 KB) and BeagleBone (256 MB), we are assuming that due to memory constraints, multimedia applications could not be implemented in TelosB nodes and memory will not be a limiting factor in BeagleBone nodes, since processing or transmission rate will limit long before these applications rather than memory. Summing up, the requirements vector for CTA and ATC applications are respectively $r_j = \left\{\text{20 kbps, 10 KB}<m_j<<\text{256 MB,17.64 MIPS}\right\}$
$r_j = \left\{\text{12 kbps, 10KB}<m_j<<\text{256 MB, 69.23 MIPS}\right\}$.

\subsection{Simulation Environment}

Sensor nodes are deployed in a $200 \times 200$ m scenario. We consider a default sensing range of $R_i^{s} = 30$ m for all of the sensors \cite{Chen:2007}. A two-ray ground path loss model with $\gamma = 4$ and $g_0 = 8.1 \cdot 10^{-3}$ \cite{Suh07} is considered. $P_{max}$ is set to 0 dBm and the receiver sensitivity $\alpha$ is $-92$ dBm \cite{cc2420}, which implies a maximum transmission range $R^T_{max}$ of 59 m. Analogously the interference sensitivity is $-104$ dBm, which implies a maximum interference range $R^I_{max}$ of 118 m.

\begin{figure*}
	\centering
	{
		\subfigure[]{\includegraphics[width=3in]{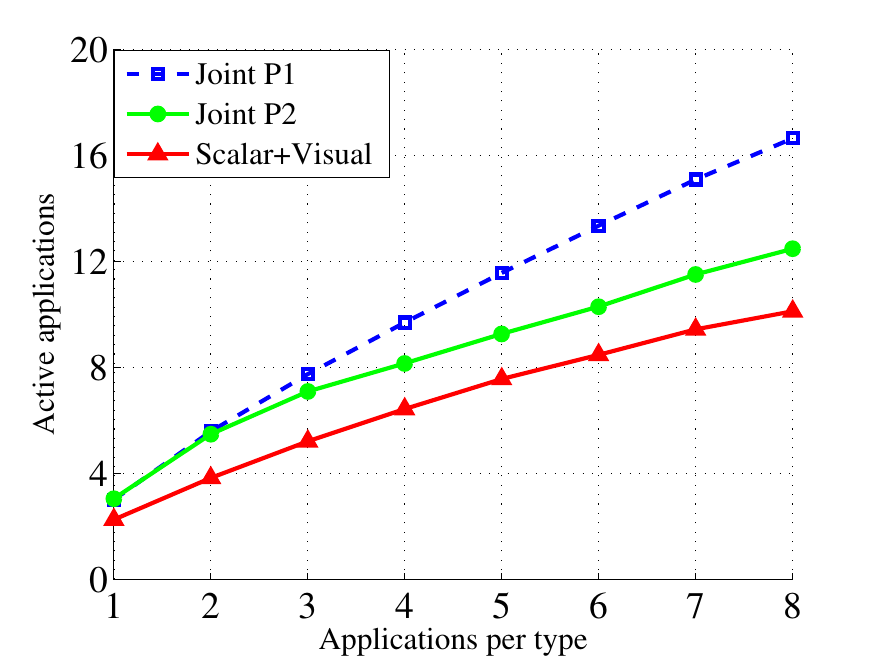}
			\label{fig:totApps}}
		\subfigure[]{\includegraphics[width=3in]{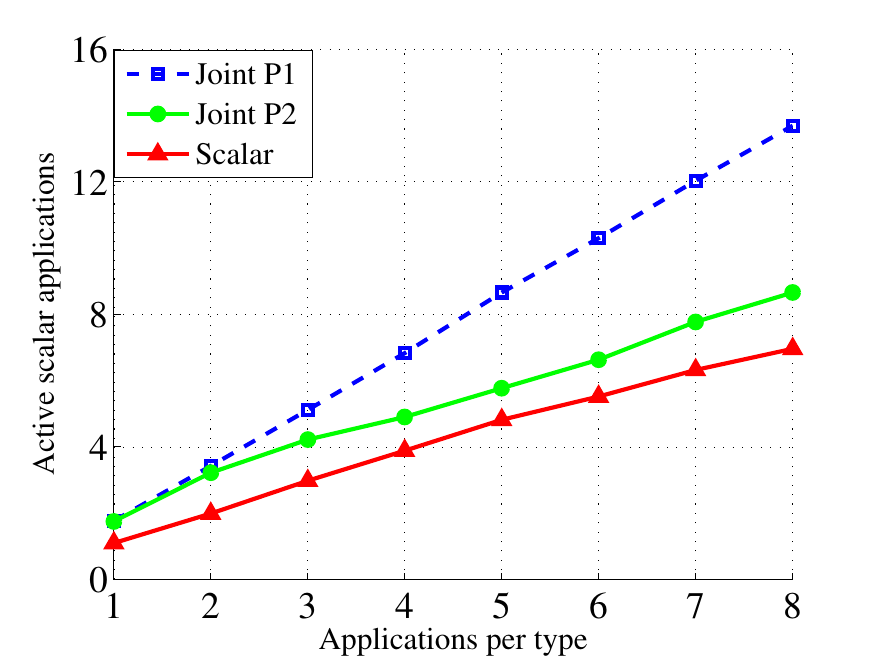}
			\label{fig:sApps}}
		\subfigure[]{\includegraphics[width=3in]{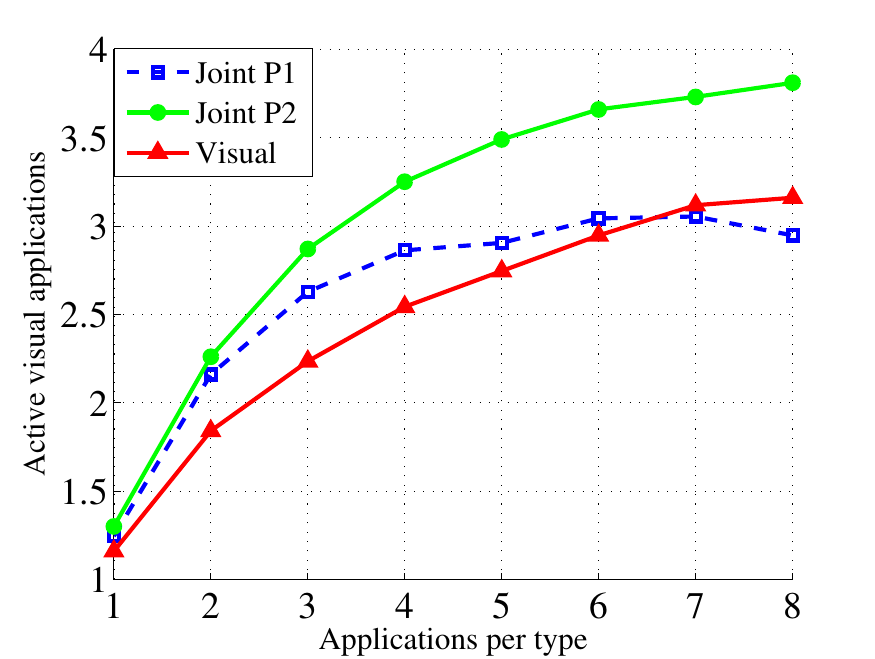}
			\label{fig:mApps}}
		
		\caption{Number of active applications vs. offered applications per type. a) Total b) Scalar c) Visual}
		\label{fig:Apps}
	}
	\end{figure*}
	
	\begin{figure*}
		\centering
		{
			\subfigure[]{\includegraphics[width=3in]{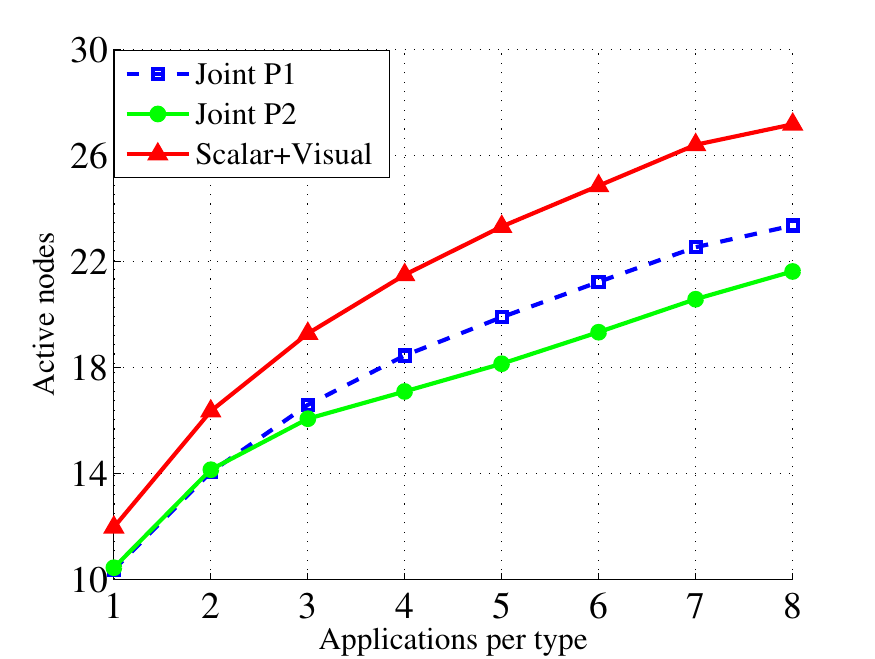}
				\label{fig:totNodes}}
			\subfigure[]{\includegraphics[width=3in]{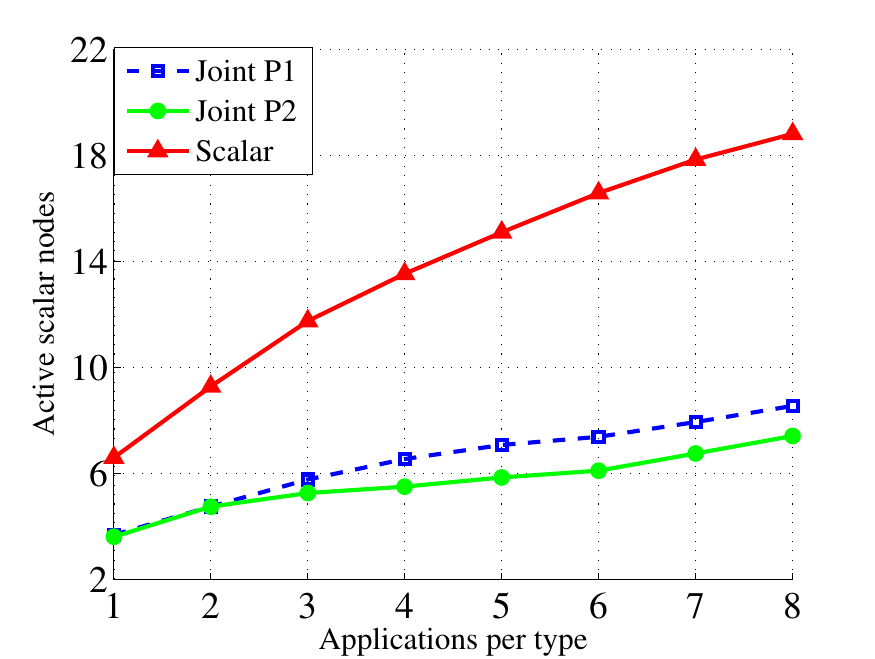}
				\label{fig:sNodes}}
			\subfigure[]{\includegraphics[width=3in]{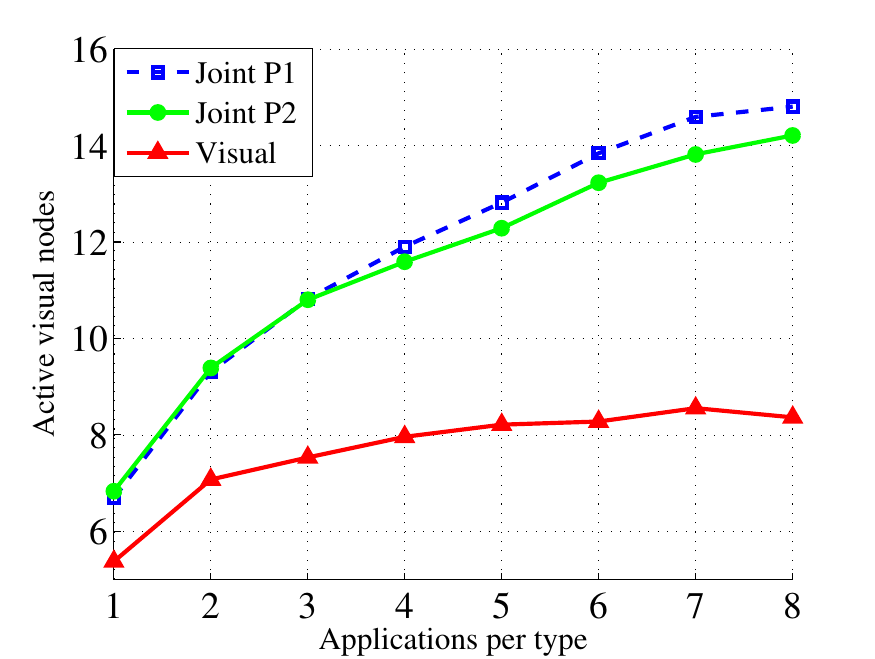}
				\label{fig:mNodes}}
			
			\caption{Number of active nodes vs. offered applications per type. a) Total b) Scalar c) Visual}
			\label{fig:Nodes}
			
		}
	\end{figure*}

\subsection{Benefits of virtualization}
\label{benefits}

As a reference example to evaluate the validity of the model and the benefits of virtualization, we have considered a scenario with two different and overlapped WSNs: a \textit{scalar} network, formed by 36 TelosB nodes and oriented to scalar applications (temperature and light monitoring), and a \textit{multimedia} network, formed by 36 BeagleBone nodes and oriented to visual applications (CTA and ATC). The number of test points is 5 for the scalar applications and 3 for the visual ones. We assume that each sensor is able to cover $N_{i,j} = 1$ test points of the same application and that each network has a sink node (one of the 36 nodes). The minimum lifetime for the virtual sensor network is $L=1$ day.

Figs.~\ref{fig:Apps} and \ref{fig:Nodes} show the performance of both networks in terms of the number of active applications and the number of active nodes when the WSNs work isolated and also when the 72 nodes cooperate as a single network that gives support to all the applications. For each point in the curves, the same number of applications of each type (temperature, light, CTA and ATC) is generated, which is the value shown in the \textit{x}-axis. For example, a 2 value in the \textit{x}-axis represents a scenario where 2 temperature, 2 light, 2 CTA and 2 ATC applications try to be deployed. ``Scalar'' refers to the single \textit{scalar} network, ``Visual'' to the single \textit{visual} network and ``Joint'' to the cooperative whole network. In this latter case, two different preference vectors $Q$ are included: P1 (all the applications have the same preference), and P2 (the preference vector depends on the application parameters). Since the main limiting factor for visual applications with regard to scalar ones is the bandwidth, preference values are approximately adjusted according to the demanded bandwidth: preference for scalar applications is 1 (they need 2.5 or 5 kbps $=0.5$ or 1 kbps per test point $\times$ 5 test points); preference for ATC applications is 8 (they require 36 kbps $=$ 12 kbps $\times$ 3 test points); and preference for CTA applications is 12 (they 60 kbps $=$ 20 kbps $\times$ 3 test points).

Fig.~\ref{fig:totApps} shows that the total number of active applications increases in the joint scenario, when compared to the sum of the independent networks (specially for the preference vector P1). As can be seen in Figs.~\ref{fig:sApps}~and~\ref{fig:mApps}, the main increase is due to the scalar applications. In fact, when visual applications are not prioritized (P1), the increase in the number of scalar applications eventually leads to an starvation of the visual applications (Fig.~\ref{fig:mApps}). However, with the preference vector P2 both scalar and visual applications experience a more balanced improvement.

The reasons for this improvement are different in each case: according to our measurements, the probability of a test point not being covered with 36 nodes is about 0.15. Since an active application requires all its test points to be effectively covered, the probability of a scalar application with 5 test points not being fully covered (only as topology concerns) is about 0.55. As multimedia nodes can support scalar applications as well, in the joint scenario the 72 nodes can be used to sense scalar applications, reducing these probabilities to 0.015 and 0.07 respectively. On the contrary, since scalar nodes do not support multimedia applications, this kind of gain cannot be obtained for visual applications. Nevertheless, as one of the main limitations in multimedia networks is bandwidth (specifically the bottleneck in the transmission to the sink node), the possibility of using two sink nodes in the joint scenario leads also to an improvement for this case. Since the resources consumed by scalar applications is much lower than by visual ones, prioritizing the latter (P2) in the objective function is useful to balance the amount of resources used by each of them.

Finally, Fig.~\ref{fig:totNodes} shows that the total number of active nodes when both networks works jointly is lower than when they work isolated. Additionally, as the number of active applications is also higher, the active nodes per active application are quite lower when the networks work jointly. Regarding the type of active nodes, Fig.~\ref{fig:mNodes} shows that the amount of active multimedia nodes increases for the joint scenario. The reasons for this effect are two: (1) multimedia nodes can support scalar applications in the joint scenario and (2) we have assumed that the cost for activating sensor nodes, $\delta_i$, is the same for both kind of nodes. We have set $\delta_i=0.01$ to ensure that in any case the cost of activating the 72 nodes is higher than the revenue of activating an application.

Once the main benefits of virtualization in the network performance have been shown, next sections present the impact of varying the most relevant model parameters. We take as basis the scenario with 36 scalar nodes and 36 visual nodes, preference vector P2, and 1 sink for each type of application (reference scenario).

\subsection{Number of scalar and multimedia nodes}
\label{numbernodes}

\begin{figure}
\centering
{
		\subfigure[]{\includegraphics[width=3in]{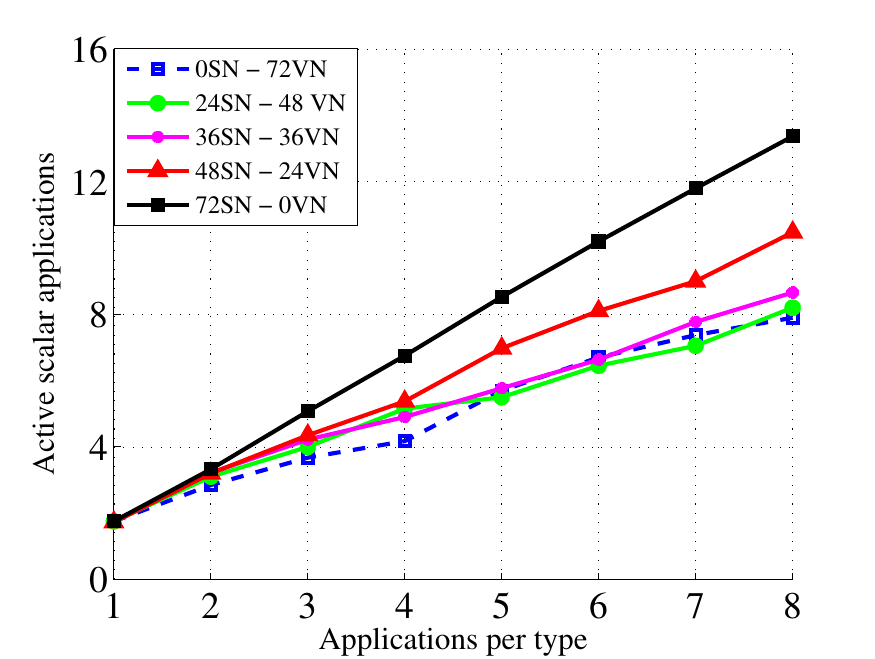}
			\label{fig:2_sApps}}
		\subfigure[]{\includegraphics[width=3in]{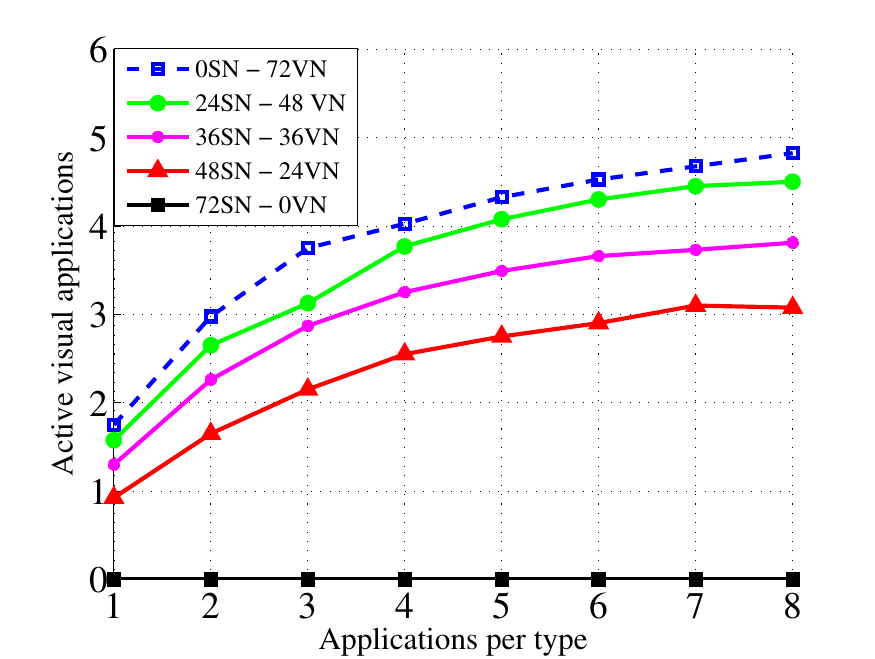}
			\label{fig:2_mApps}}
		
		\caption{Number of active applications vs. offered applications per type varying the type of nodes. a) Scalar b) Visual}
		\label{fig:2_Apps}
	}
\end{figure}

\begin{figure}
		\centering
		{
			\subfigure[]{\includegraphics[width=3in]{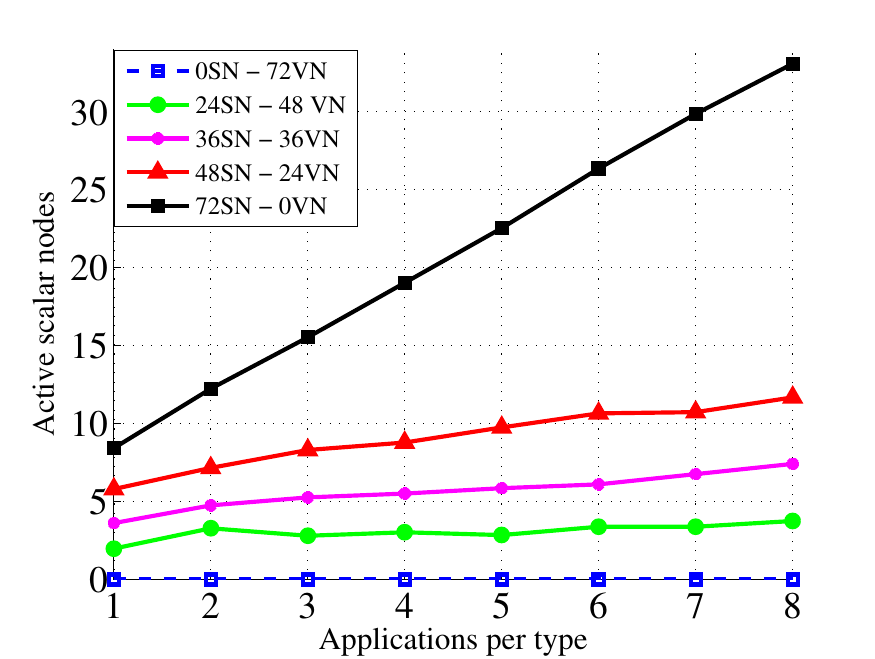}
				\label{fig:2_sNodes}}
			\subfigure[]{\includegraphics[width=3in]{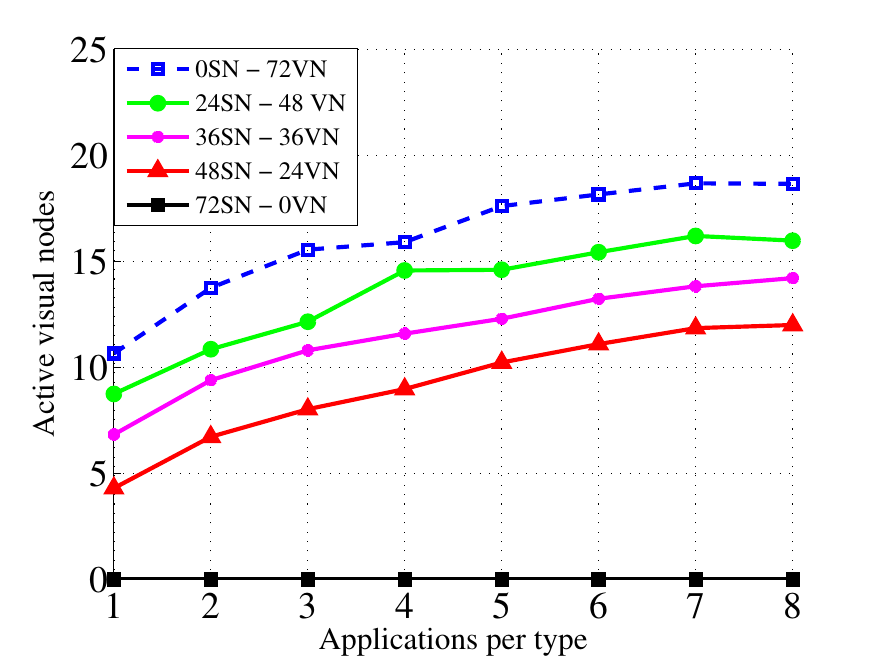}
				\label{fig:2_mNodes}}
			
			\caption{Number of active nodes vs. offered applications per type varying the type of nodes. a) Scalar b) Visual}
			\label{fig:2_Nodes}
			
		}
	\end{figure}

First, we vary the number of scalar and multimedia motes, maintaining the total number of nodes to 72. Fig.~\ref{fig:2_Apps} and Fig.~\ref{fig:2_Nodes} show the number of active applications and the number of active nodes respectively. As can be seen, the probability of a visual application being fully covered increases with the number of multimedia motes since visual applications can only be deployed on multimedia motes. Therefore, the number of active visual applications grows from 0 (when there are not multimedia motes) until its maximum value (when the 72 motes are multimedia) (Fig.~\ref{fig:2_mApps}). However, the growth rate of active visual applications decreases as the number of multimedia motes grows due to the bandwidth bottleneck at the sinks. The number of active scalar applications (fig.~\ref{fig:2_sApps}) is not impacted by the number of scalar motes, since they can be deployed on any mote and are less limited by bandwidth constraints.

	\begin{figure}
	\centering
	{
		\subfigure[]{\includegraphics[width=3in]{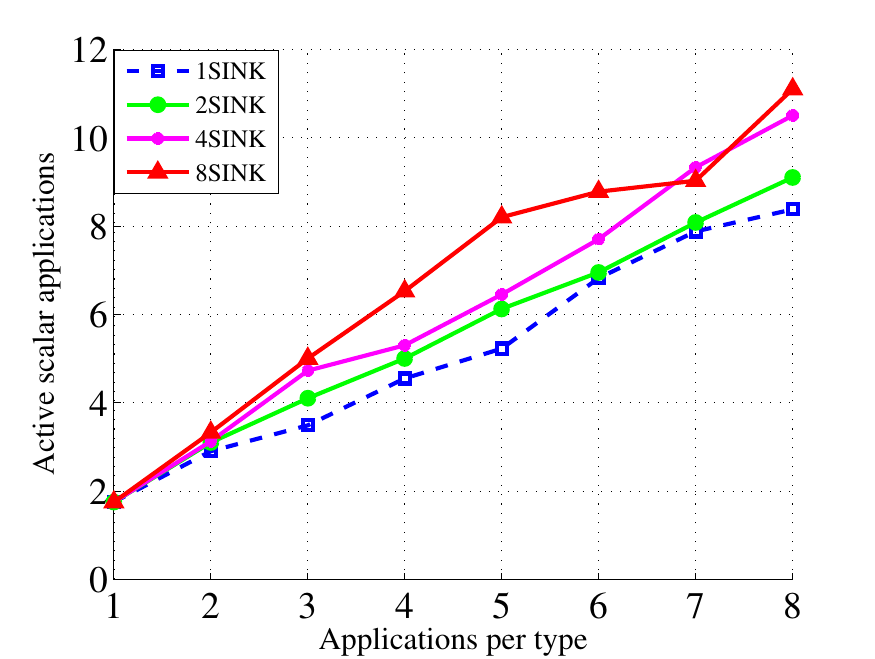}
			\label{fig:new3_sApps}}
		\subfigure[]{\includegraphics[width=3in]{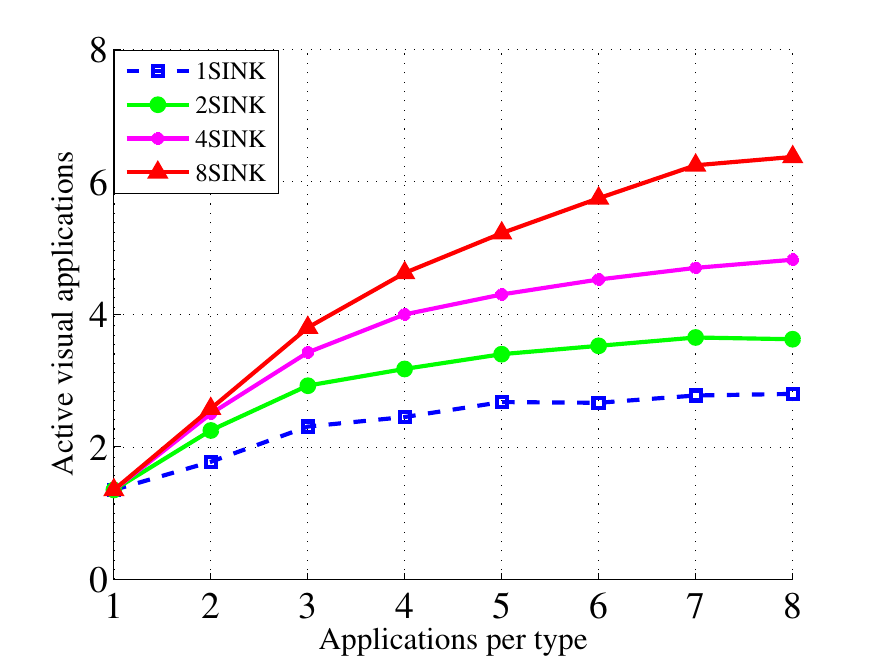}
			\label{fig:new3_mApps}}
		
		\caption{Number of active applications vs. offered applications per type varying the number of sinks. a) Scalar b) Visual}
		\label{fig:new3_Apps}
	}
	\end{figure}
	
	\begin{figure}
		\centering
		{
			\subfigure[]{\includegraphics[width=3in]{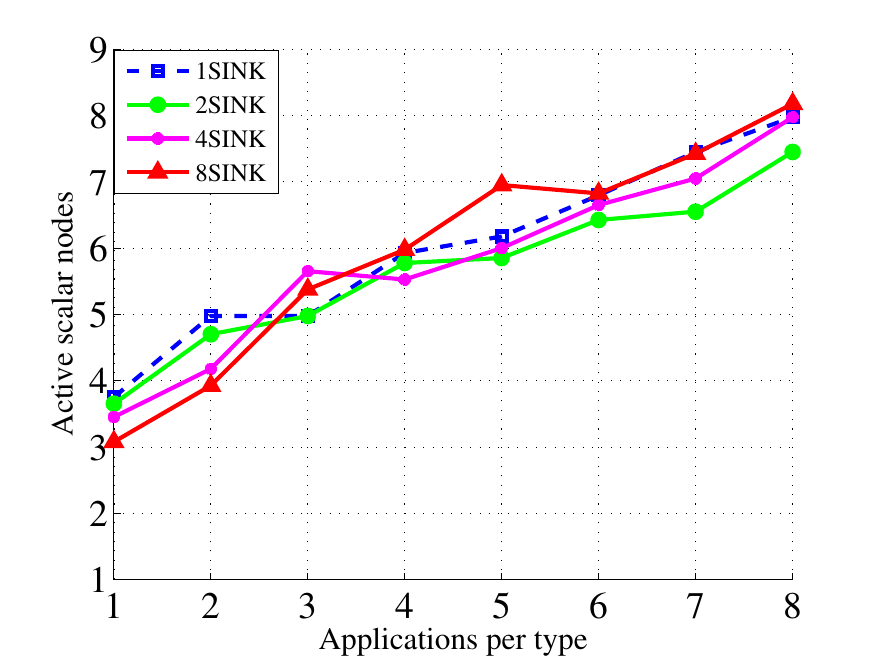}
				\label{fig:new3_sNodes}}
			\subfigure[]{\includegraphics[width=3in]{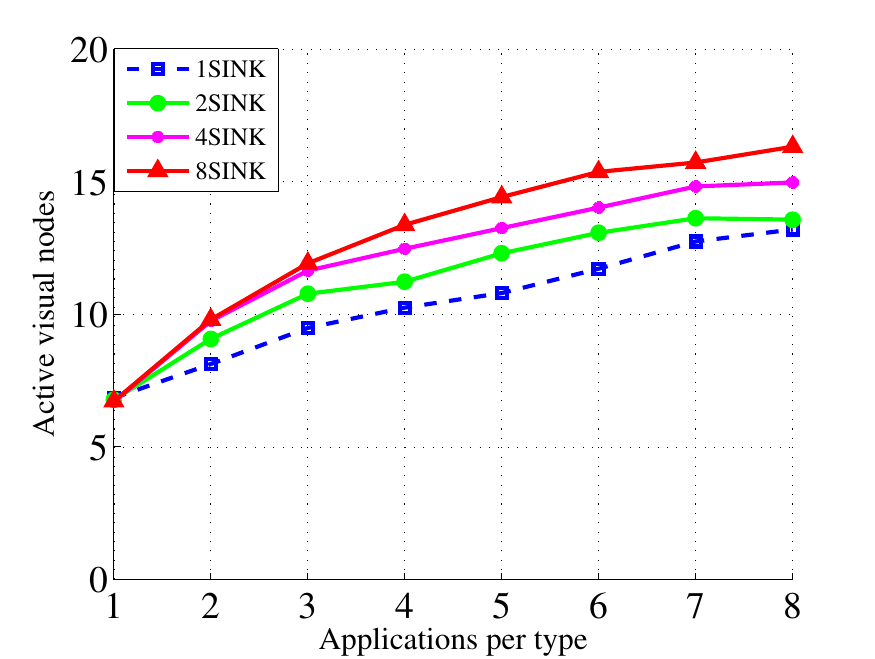}
				\label{fig:new3_mNodes}}
			
			\caption{Number of active nodes vs. offered applications per type varying the number of sinks. a) Scalar b) Visual}
			\label{fig:new3_Nodes}
			
		}
	\end{figure}

\begin{figure}
	\centering
	{
		\subfigure[]{\includegraphics[width=3in]{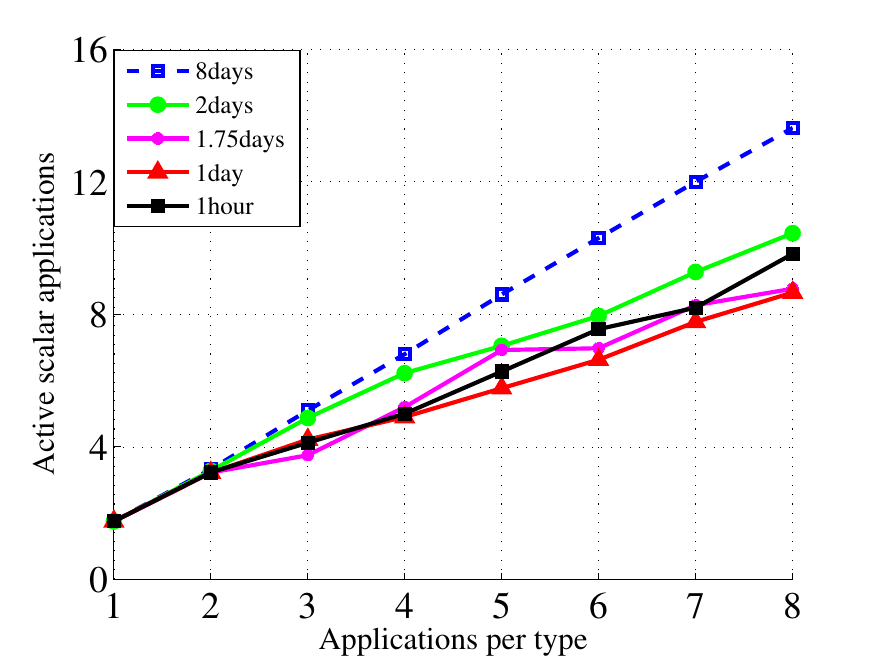}
			\label{fig:5_sApps}}
		\subfigure[]{\includegraphics[width=3in]{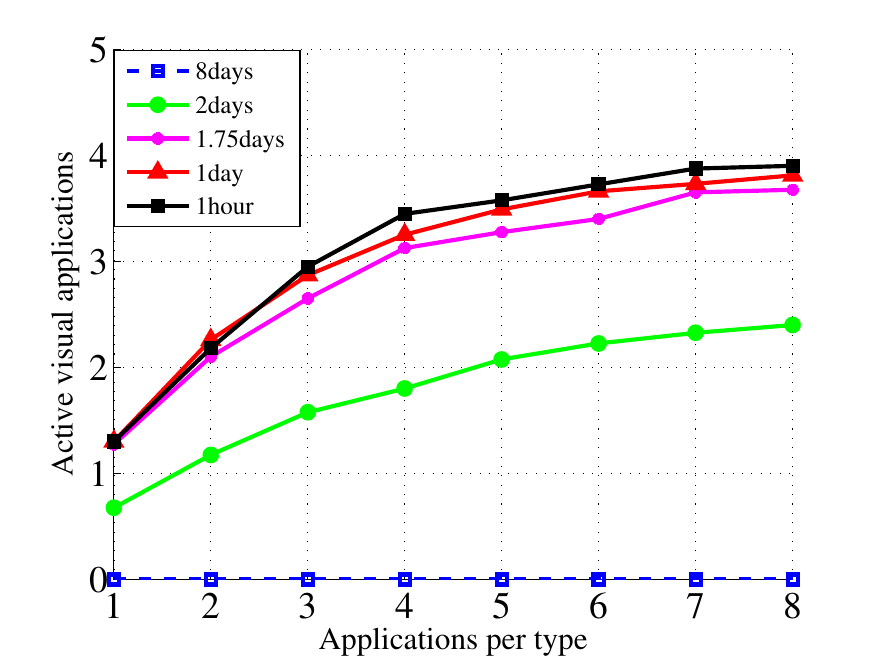}
			\label{fig:5_mApps}}
		
		\caption{Number of active applications vs. offered applications per type varying the network lifetime. a) Scalar b) Visual}
		\label{fig:5_Apps}
	}
	\end{figure}
	
	\begin{figure}
		\centering
		{
			\subfigure[]{\includegraphics[width=3in]{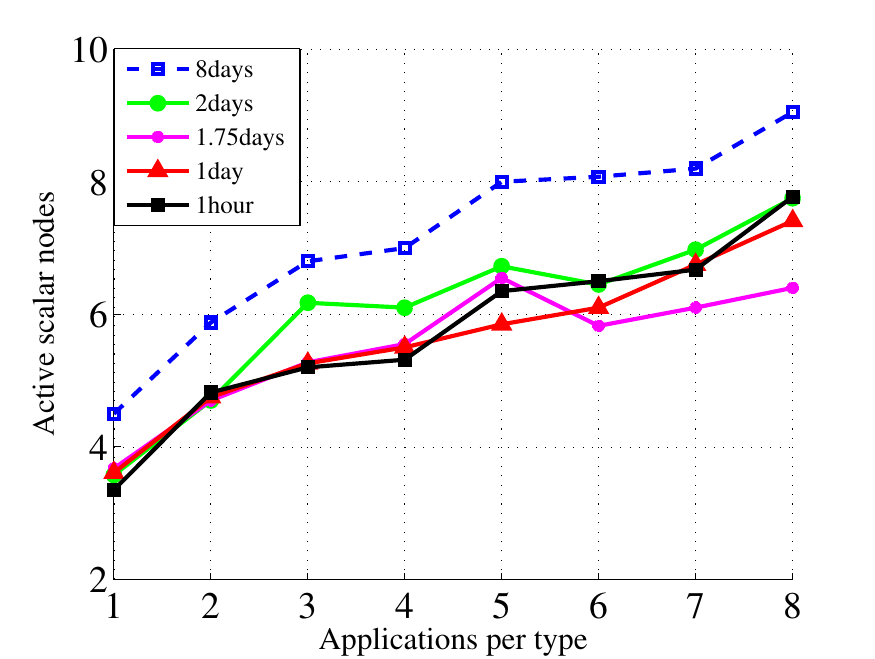}
				\label{fig:5_sNodes}}
			\subfigure[]{\includegraphics[width=3in]{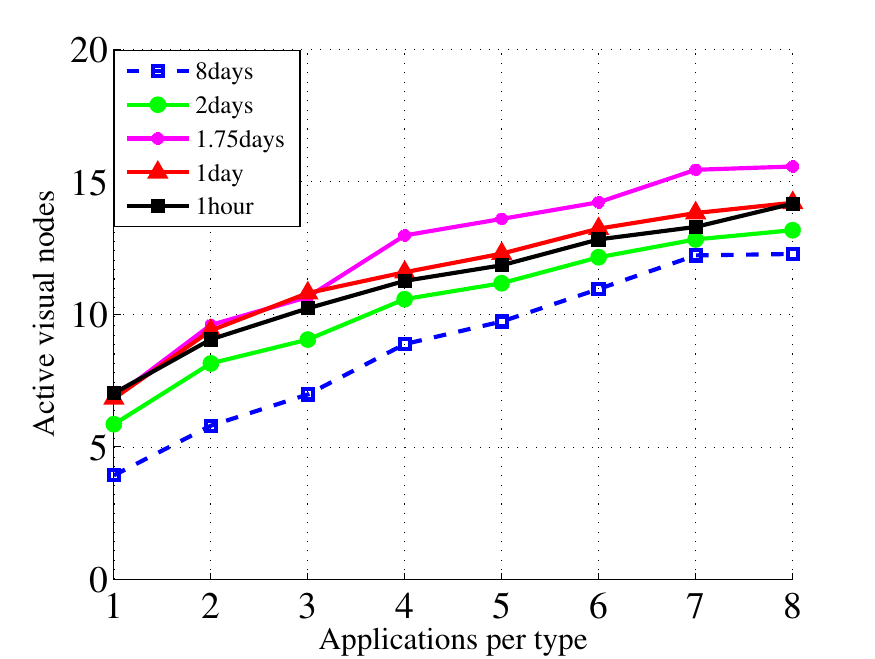}
				\label{fig:5_mNodes}}
			
			\caption{Number of active nodes vs. offered applications per type varying the network lifetime. a) Scalar b) Visual}
			\label{fig:5_Nodes}
			
		}
	\end{figure}

\subsection{Number of sinks}
\label{numbersinks}
	
Since the number of active multimedia applications depends on the bottleneck access to the sink, we analyze next the impact of the number of sinks in the reference scenario. Figs.~\ref{fig:new3_Apps} and \ref{fig:new3_Nodes}, that show the network performance as a function of the number of sinks, confirm that assumption. As can be seen, there is a great improvement in the number of active visual applications when the number of sinks is increased. Additionally, there is also an increase in the number of active scalar applications, which is not so relevant due to their lower bandwidth requirements. Consequently, the number of active scalar nodes (Fig.~\ref{fig:new3_sNodes}) does not depend on the number of sinks, whereas the number of active multimedia nodes (Fig.~\ref{fig:new3_mNodes}) increases with the number of sinks since the number of active visual applications grows.

\subsection{Lifetime}
\label{lifetime}

In the reference scenario, the minimum lifetime of the virtual sensor network is set to $L=1$ day. Next, we vary this parameter from $L=1$ hour to $L=8$ days. The results are shown in Figs.~\ref{fig:5_Apps}~and~\ref{fig:5_Nodes}. Fig.~\ref{fig:5_mApps} shows that with $L=8$ days, visual applications cannot be activated since multimedia motes do not have enough energy to support any visual application. Logically, multimedia motes (fig.~\ref{fig:5_mNodes}) are still activated because they can be used by the scalar applications. In addition, there is a remarkable decrease in the number of active visual applications from $L=1.75$ to $L=2$ days. The reason is that ATC visual applications, which demand more energy, cannot be activated with $L=2$ days and the only active visual applications are the CTA ones. It is also interesting to observe that from $L=1$ to $L=1.75$ days there is a slight decrease in the number of active visual applications whereas the number of active multimedia motes rises. This is due to the fact that nodes that were simultaneously sensing several test points when $L=1$, do not have now enough energy when $L=1.75$ and therefore, additional nodes must be activated.

	\begin{figure}
	\centering
	{
		\subfigure[]{\includegraphics[width=3in]{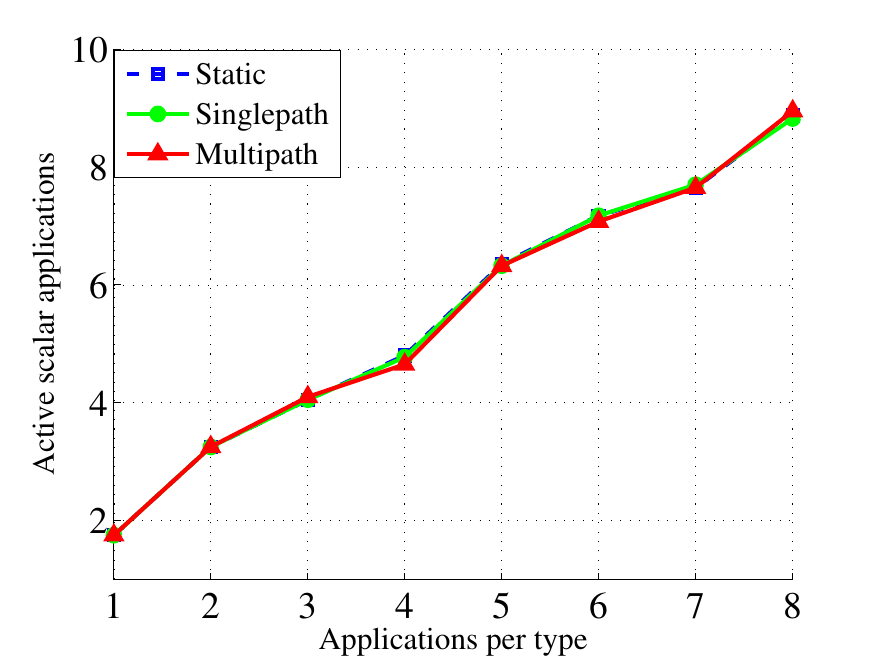}
			\label{fig:new4_sApps}}
		\subfigure[]{\includegraphics[width=3in]{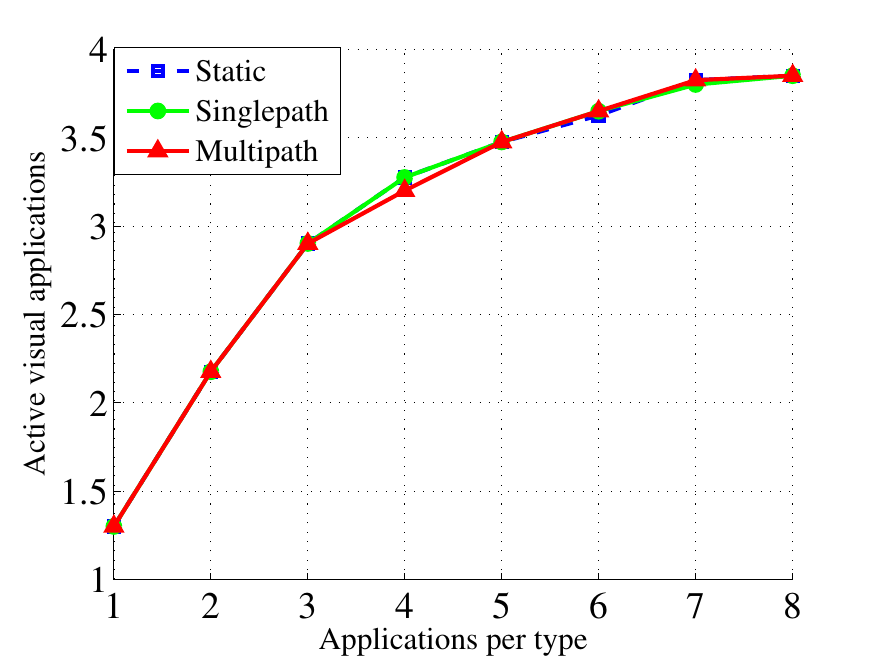}
			\label{fig:new4_mApps}}
		
		\caption{Number of active applications vs. offered applications per type varying the routing schemes. $P_{max} = 0dBm$. a) Scalar b) Visual}
		\label{fig:new4_Apps}
	}
	\end{figure}
	
	\begin{figure}
		\centering
		{
			\subfigure[]{\includegraphics[width=3in]{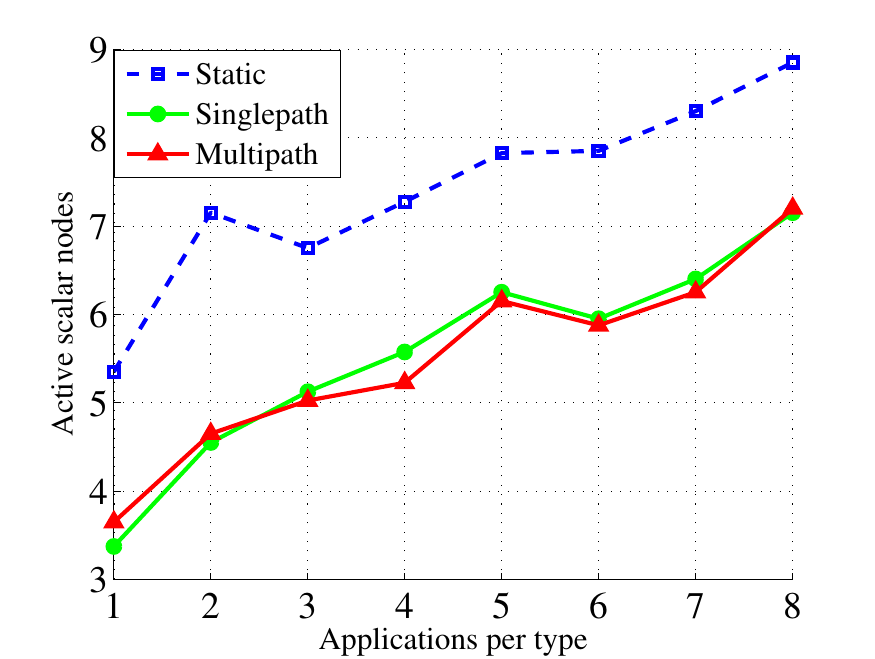}
				\label{fig:new4_sNodes}}
			\subfigure[]{\includegraphics[width=3in]{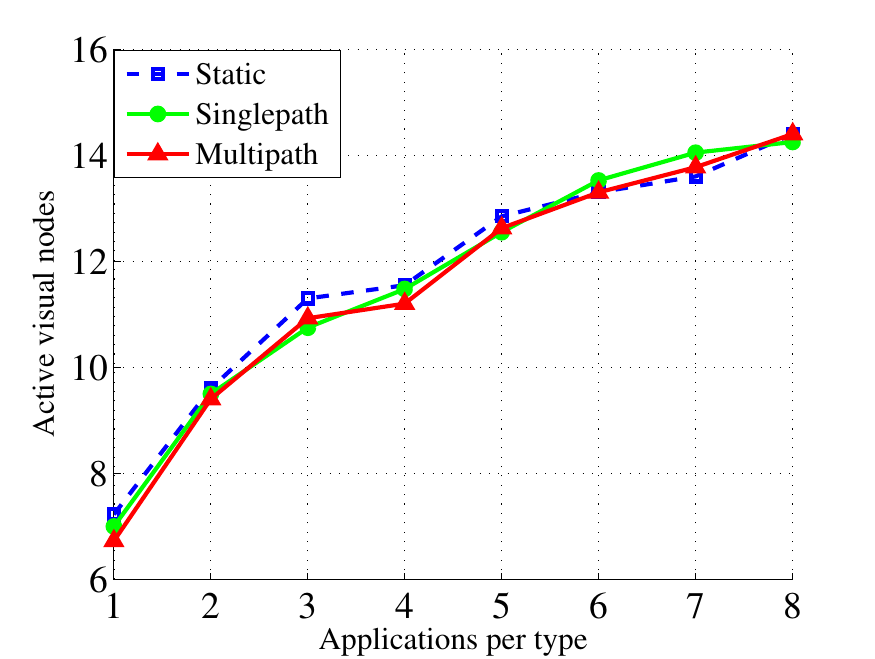}
				\label{fig:new4_mNodes}}
			
			\caption{Number of active nodes vs. offered applications per type varying the routing schemes. $P_{max} = 0dBm$. a) Scalar b) Visual}
			\label{fig:new4_Nodes}
			
		}
	\end{figure}

		\begin{figure}
	\centering
	{
		\subfigure[]{\includegraphics[width=3in]{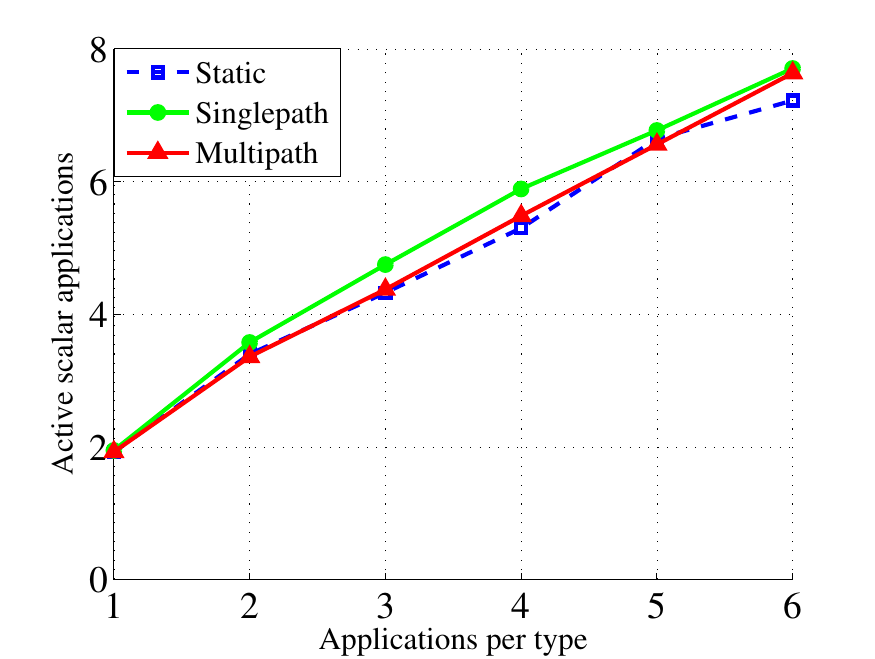}
			\label{fig:6_sApps}}
		\subfigure[]{\includegraphics[width=3in]{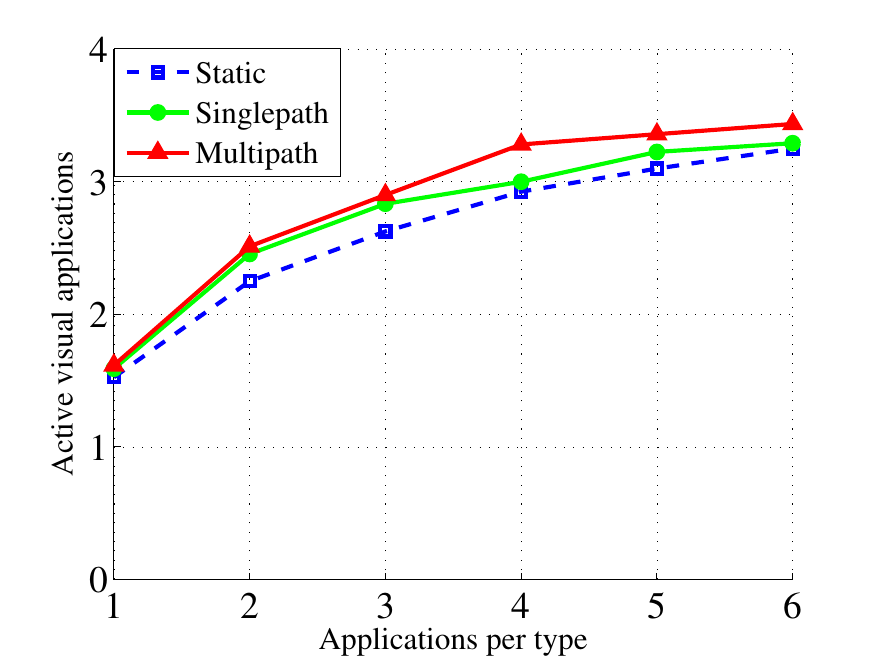}
			\label{fig:6_mApps}}
		
		\caption{Number of active applications vs. offered applications per type varying the routing schemes. $P_{max} = -10dBm$. a) Scalar b) Visual}
		\label{fig:6_Apps}
	}
	\end{figure}
	
	\begin{figure}
		\centering
		{
			\subfigure[]{\includegraphics[width=3in]{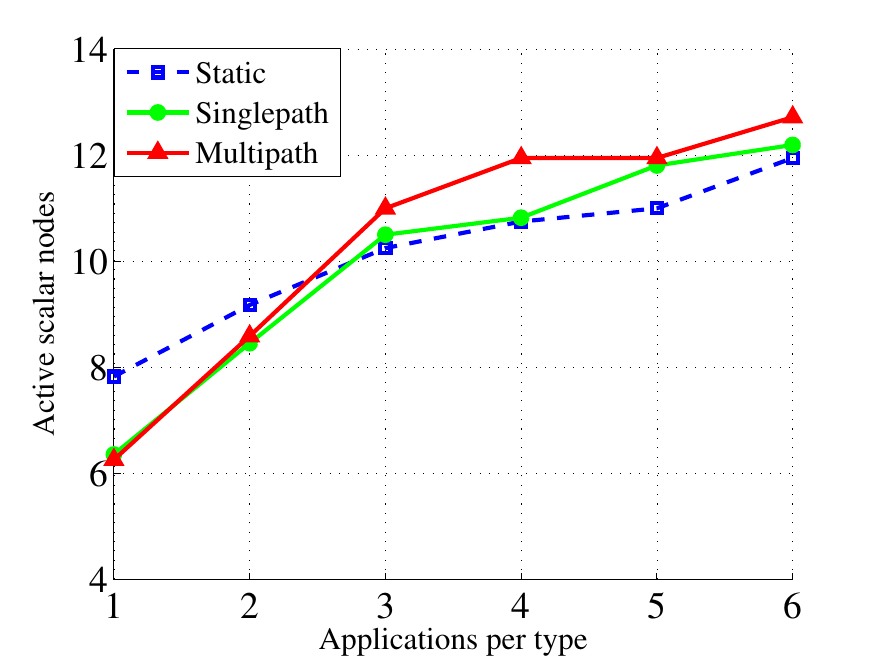}
				\label{fig:6_sNodes}}
			\subfigure[]{\includegraphics[width=3in]{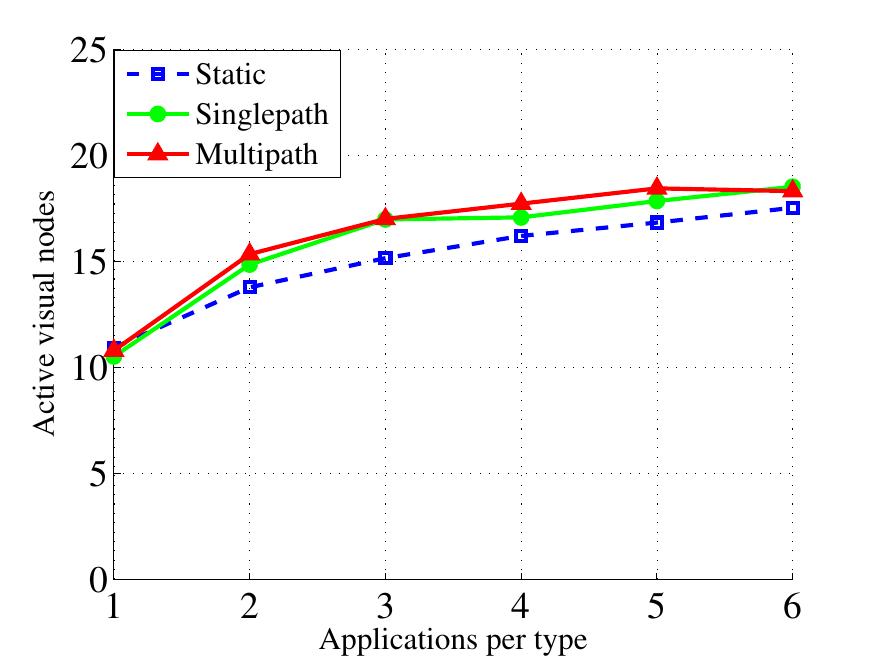}
				\label{fig:6_mNodes}}
			
			\caption{Number of active nodes vs. offered applications per type varying the routing schemes. $P_{max} = -10dBm$. a) Scalar b) Visual}
			\label{fig:6_Nodes}
			
		}
	\end{figure}

\subsection{Type of routing}
\label{routing}

Finally, the three different types of routing described in section~\ref{routing-const} are analyzed: \textit{multipath}, \textit{singlepath} and \textit{static routing}. Figs.~\ref{fig:new4_Apps} and~\ref{fig:new4_Nodes} show the number of active applications and the number of active nodes for the three different routing schemes in the reference scenario. As can be seen, the impact of the routing in the results seems very limited. Nevertheless, the reason behind this is that most of the routes only have one or two hops (with $P_{max} = 0$ dBm, the maximum transmission range is 59 m). To increase the number of hops of the routes, we decrease $P_{max}$ to $-10$ dBm (maximum transmission range of 33 m). Figs.~\ref{fig:6_Apps} and~\ref{fig:6_Nodes} show these new results. In this case, it can be observed that the \textit{singlepath} routing achieves a performance very close to the upper bound provided by the ideal non-constrained \textit{multipath} routing. In addition, the much simpler \textit{static} routing is also close to the \textit{singlepath}, which suggests its utilization in the proposed heuristic algorithm, whose performance is shown next. It can also be noted that the number of active nodes (Fig.~\ref{fig:6_Nodes}) rises due to the increase in the number of hops.

\begin{figure*}
	\centering
	{
		\subfigure[]{\includegraphics[width=3in]{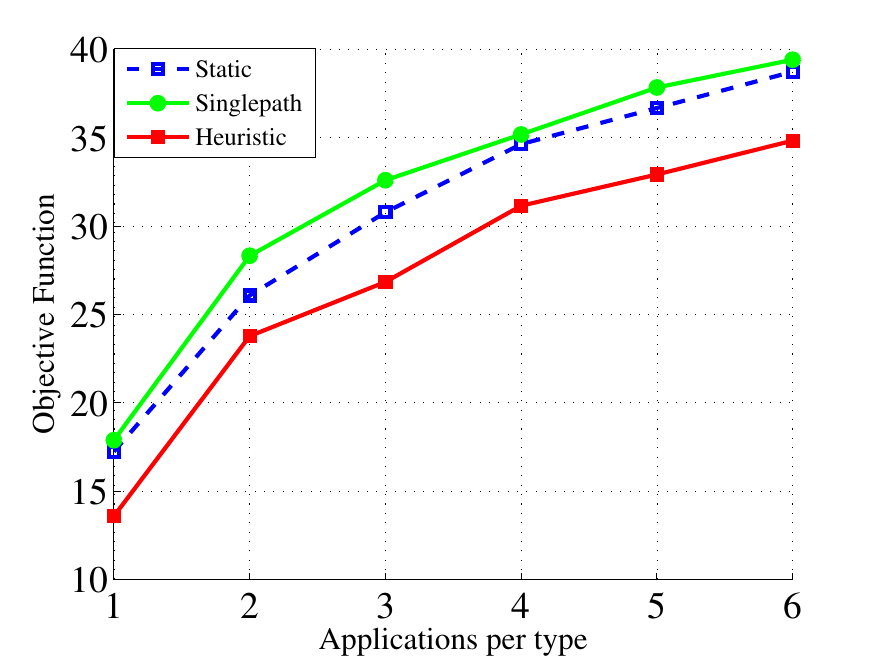}
			\label{fig:Heu_Profit}}
		\subfigure[]{\includegraphics[width=3in]{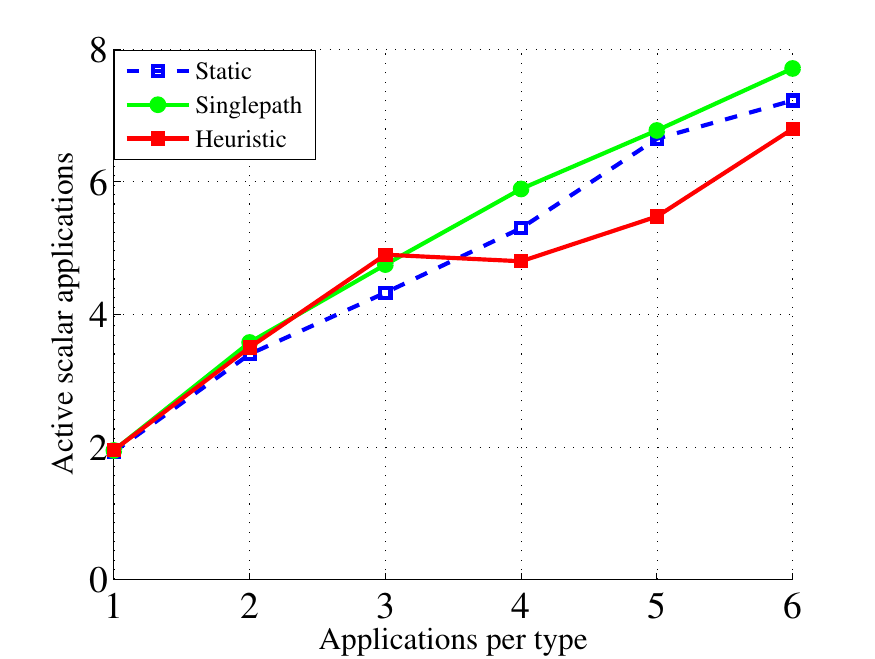}
			\label{fig:Heu_sApps}}
		\subfigure[]{\includegraphics[width=3in]{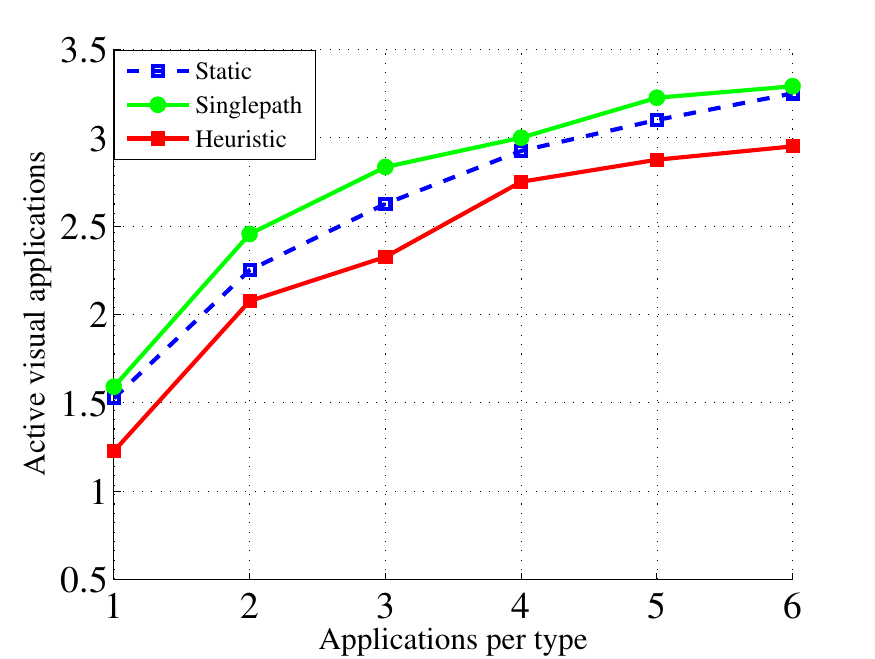}
			\label{fig:Heu_mApps}}
		
		\caption{Performance of heuristic algorithm vs the optimum scheme with \textit{singlepath} and \textit{static} routing. Number of active applications vs. offered applications per type. a) Objective function b) Active scalar applications c) Active visual applications}
		\label{fig:Heu}
	}
	\end{figure*}

\subsection{Performance of heuristic algorithm}
\label{Heuristic-results}

In this section, the heuristic algorithm is evaluated in the reference scenario with $P_{max} = -10$ dBm to see the impact of the use of the static routing in the heuristic. Fig.~\ref{fig:Heu} compares the performance of the proposed heuristic algorithm with the optimum solutions achieved with the \textit{singlepath} routing and with the \textit{static} routing. As can be seen in Fig.~\ref{fig:Heu_Profit}, the degradation of the heuristic algorithm is about a 10\% from the optimum value. A similar decrease is observed in the number of active scalar and visual applications (Figs.~\ref{fig:Heu_sApps}~and~\ref{fig:Heu_mApps}). These results suggest the potential of this algorithm as a centralized resource allocation tool for virtual sensor networks.

\section{Conclusion}
\label{conclusion}
In this paper we have analyzed a virtual sensor network where different kinds of applications and sensor nodes coexist and cooperate. We have formulated mathematically the optimization problem of maximizing the overall revenue out of the application deployment process while minimizing the cost related to activating sensor nodes and we have analyzed its computational complexity. Constraints regarding sensor nodes capabilities (memory, computation, energy) and network limitations (topology, shared bandwidth) have been included. A heuristic algorithm has been proposed to reduce the computation time of the resource allocation pattern. Realistic parameters for both the sensor nodes and the supported applications have been considered in the evaluation of the model. Simulation results are further derived to assess the potential performance gains that the resource reuse achieved by virtualization can obtain: coverage enhancements, since there is a higher density of sensor nodes capable of supporting a given application, and capacity increase, due to the possibility of reusing several sink nodes to reduce congestion on bottleneck links.

\section*{Acknowledgment}

This work has been supported by the Spanish Government through the grants TEC2011-23037 and TEC2014-52969-R from the Ministerio de Ciencia e Innovaci\'on (MICINN), Gobierno de Arag\'on (research group T98), the European Social Fund (ESF) and Centro Universitario de la Defensa through project CUD2013-05.

This work has also been partially supported by the Italian Ministry for Education, University and Research (MIUR) through the national cluster project SHELL, Smart Living technologies (grant number: CTN01 00128 111357).

\section*{References}

\bibliography{mybibfile}

\end{document}